\documentclass[prl,aps,twocolumn,floats,nofootinbib,showpacs]{revtex4}
\usepackage{amsmath,amssymb,graphicx,psfrag}

\begin{document}

\renewcommand{\ni}{{\noindent}}
\newcommand{\dprime}{{\prime\prime}}
\newcommand{\be}{\begin{equation}}
\newcommand{\ee}{\end{equation}}
\newcommand{\bea}{\begin{eqnarray}} 
\newcommand{\eea}{\end{eqnarray}}
\newcommand{\la}{\langle}
\newcommand{\ra}{\rangle} 
\newcommand{\dg}{\dagger}
\newcommand\lbs{\left[}
\newcommand\rbs{\right]}
\newcommand\lbr{\left(}
\newcommand\rbr{\right)}
\newcommand\f{\frac}
\newcommand\e{\epsilon}
\newcommand\ua{\uparrow}
\newcommand\da{\downarrow}
\newcommand{\bitato}{0^{\beta}}
\def \d {^\dagger}
\def \being {\begin{equation*}}
\def \ening {\end{equation*}}
\def \mk {\mathbf{k}}

\title{Phase Diagram of the Half-Filled Ionic Hubbard Model}
\author{Soumen Bag$^{1}$, Arti Garg$^{2}$, and H. R. Krishnamurthy$^{1,3}$} 
\affiliation{$^{1}$ Department of Physics, Indian Institute of Science, Bangalore 560 012, India \\
$^{2}$Condensed Matter Physics Division, Saha Institute of Nuclear Physics, 1/AF Bidhannagar, Kolkata 700 064, India\\
$^{3}$Physics Department, University of California Santa Cruz, California 95064, USA}
\vspace{0.2cm}
\begin{abstract}
\vspace{0.3cm}
We study the phase diagram of the ionic Hubbard model (IHM) at half-filling on a Bethe lattice of infinite connectivity using dynamical mean field theory (DMFT), with two impurity solvers, namely, iterated perturbation theory (IPT) and continuous time quantum Monte Carlo (CTQMC). 
The physics of the IHM is governed by the competition between the staggered ionic potential $\Delta$ and the on-site Hubbard $U$. We find that for a finite $\Delta$ and at zero temperature,  long range anti-ferromagnetic (AFM) order sets in beyond a threshold $U=U_{AF}$ via a first order phase transition. For $U$ smaller than $U_{AF}$ the system is a correlated band insulator. Both the methods show a clear evidence for  a quantum transition to a half-metal phase just after the AFM order is turned on, followed by the formation of an AFM insulator on further increasing $U$.  We show that the results obtained within both the methods have good qualitative and quantitative consistency in the intermediate to strong coupling regime at zero temperature as well as at finite temperature. On increasing the temperature, the AFM order is lost via a first order phase transition at a transition temperature $T_{AF}(U, \Delta)$ (or, equivalently, on decreasing $U$ below $U_{AF}(T, \Delta)$), within both the methods, for weak to intermediate values of $U/t$. In the strongly correlated regime,
where the effective low energy Hamiltonian is the Heisenberg model, IPT is unable to capture the thermal (Neel) transition from the AFM phase to the paramagnetic phase, but the CTQMC does. At a finite temperature $T$, DMFT+CTQMC shows a second phase transition (not seen within DMFT+IPT) on increasing $U$ beyond $U_{AF}$. At $U_N > U_{AF}$, when the Neel temperature $T_N$ for  the effective Heisenberg model becomes lower than $T$, the AFM order is lost via a second order transition. For $U \gg \Delta$, $T_N \sim t^2/U(1-x^2)$ where $x=2\Delta/U$ and thus $T_N$ increases with increase in $\Delta/U$. In the 3-dimensional parameter space of $(U/t, T/t ~ and~ \Delta/t)$, as $T$ increases, the surface of first order transition at $U_{AF}(T, \Delta)$ and that of the second order transition at $U_N(T, \Delta)$ approach each other, shrinking the range over which the AFM order is stable. There is a line of tricritical points that separates the  surfaces of first and second order phase transitions.  
\vspace{0.1cm}
\typeout{polish abstract}
\end{abstract} 
\maketitle
\section{Introduction}
The Hubbard model is a paradigm for studying electron correlation effects in metallic systems in condensed matter physics. It has played an important role in understanding how electron electron interactions can give rise to many interesting phases, including insulating, magnetic and super-conducting phases. It is worthwhile to explore whether there are interesting correlation effects in systems that would be band insulators in the absence of electron-electron interactions. Perhaps the simplest model in which one can carry out this exploration is an extension of the Hubbard model, known as the Ionic Hubbard model (IHM), with a staggered on-site ``ionic'' potential $\Delta$ added in. In the recent past the IHM has been studied in various dimensions by a variety of numerical and analytical tools~\cite{1d-1,1d-2,1d-3,Jabben,AG1,qmc_ihm1,qmc_ihm2,cdmft_ihm,hartmann,kampf,Hoang,AG2,rajdeep}. In one-dimension~\cite{1d-1,1d-2,1d-3} it has been shown to have a spontaneously dimerized phase, in the intermediate coupling regime, which separates the weakly coupled band insulator from the strong coupling Mott insulator. 
In higher dimensions ($d > 1$), this model has been studied by many groups using the dynamical mean field theory (DMFT)~\cite{Jabben,AG1,hartmann,kampf,AG2,rajdeep}, determinantal quantum Monte carlo~\cite{qmc_ihm1,qmc_ihm2}, cluster DMFT~\cite{cdmft_ihm} and coherent potential approximation~\cite{Hoang}.  The solution of the DMFT self consistent equations in the paramagnetic (PM) sector at half filling at zero temperature shows an intervening correlation induced metallic phase~\cite{AG1,qmc_ihm1,qmc_ihm2,hartmann,Hoang} for intermediate values of the interaction $U$. When one allows for spontaneous spin symmetry breaking the transition from paramagnetic band insulator (PM BI) to AFM insulator generally preempts the formation of the para-metallic phase~\cite{kampf,cdmft_ihm}, except, as shown in a recent paper coauthored by two of us~\cite{AG2}  using DMFT with Iterated Perturbation Theory (IPT) as the impurity solver, for a sliver of a half-metallic AFM phase. Upon doping one gets a broad ferrimangetic half-metal phase~\cite{AG2} sandwiched between a weakly correlated PM metal for small $U$ and a strongly correlated metal for large $U$.

In this paper, we provide a detailed discussion of the properties of the half-filled IHM on the Bethe lattice of infinite connectivity solved using DMFT, not only at $T=0$, specially the half-metallic AFM phase, but also at the finite temperature at a level much more extensive than explored before~\cite{rajdeep}. The DMFT equations are solved allowing the possibility of an anti-ferromagnetic (AFM) order. 
We show that at any finite $T$, the system shows two phase transitions as the Hubbard $U$ is tuned for a fixed value of the ionic potential $\Delta$. As $U$ increases, first the AFM order turns on via a first order phase transition at $U_{AF}$ followed up by a continuous transition at $U_N > U_{AF}$ at which the staggered magnetisation drops to zero continuously. As $T$ increases, $U_{AF}$ increases while $U_N$ decreases, due to enhanced thermal fluctuations, such that the range in $U$ over which the long range AFM order survives shrinks. At a certain $T_{tcp}$, we have a tri-critical point that separates the lines of the first and the second order phase transitions (for fixed $\Delta$). In the 3D parameter space of $U-\Delta-T$, there is a line of tricritical points separating surfaces of first and second order transitions. We believe that these features of the phase diagram of the IHM have not been pointed out earlier. 

Our results come from a detailed study of the IHM model using DMFT with two different impurity solvers, namely, the iterated perturbation theory (IPT) and the continuous time quantum Monte-carlo (CTQMC- implemented using TRIQS package~\cite{Triqs}), which also allows us to explore what is the interaction regime where IPT, an approximate semi-analytic impurity solver, works well. 
We show that the zero temperature phase diagram obtained within DMFT+IPT matches well (see Fig.~\ref{phase_diag}), both qualitatively and semi-quantitatively, with that obtained from the DMFT+CTQMC (working at very low temperatures). In both the methods we find that at zero temperature, and for a finite $\Delta$, the long range AFM order sets in beyond a threshold value of $U$, which we denote $U_{AF}$, via a first order phase transition. For $U$ smaller than $U_{AF}$ the system is a correlated band insulator in which the gap in the single particle density of states (DOS) reduces as $U$ is increased. Both $U_{AF}$ and the jump in the staggered magnetization at the transition point increase with $\Delta$. Both the methods show clear evidence of the formation of a half-metal phase just after the AFM order sets in, followed by the formation of an AFM insulator (AFM I) on further increasing $U$. Note that the half-metal AFM phase is missed out completely in the Hartree-Fock theory.

For weak to intermediate ($U\sim 6t$ and thus moderately strong) couplings, where the interesting metallic and half-metallic phases are realized in this model within DMFT+IPT, there is a good quantitative consistency between the  CTQMC and IPT results. However, in the limit of extremely strong coupling, where energetically it is favorable to project out the double occupancies and the ionic Hubbard model maps onto the effective Heisenberg model at half filling, one starts seeing deviation between the CTQMC and IPT results. While DMFT+CTQMC correctly captures the physics of the effective Heisenberg model (as was also shown in~\cite{rajdeep}), perhaps not surprisingly, IPT fails to do so. At any finite temperature, CTQMC shows two phase transitions as the Hubbard $U$ is tuned. First, with increasing $U$, the long range AFM order sets in via a first order jump in the staggered magnetisation $m_s$ at $U_{AF}(T)$. On increasing $U$ further, $m_s$ first increases, reaches a maximum and then starts decreasing with $U$, finally going to zero via a continuous transition at $U = U_N(T)$. This transition happens when $T$ just crosses $T_N$ where $T_N$ is the Neel temperature of the effective Heisenberg model obtained at $U=U_N$. 

Within DMFT+IPT, at any finite $T$, only the first phase transition is seen as the Hubbard $U$ is tuned. Within IPT also the AFM order sets in with a first order jump at $U_{AF}$. However, as $U$ increases further, the AFM order saturates to unity and never goes to zero again. Thus the second phase transition from the AFM insulator to paramagnetic phase at large U is not captured by IPT. 

Consistent with this, the thermal phase transition for the half filled IHM is correctly captured within IPT only for weak to intermediate $U$, but is correctly described by the CTQMC for strong correlations as well. For all values of $U > U_{AF}$, the thermal transitions to the paramagnetic phase shown by IPT are always first order. On the other hand, CTQMC shows a first order transition only for small values of $U/t$. For $U \gg 2\Delta$, the staggered magnetisation falls to zero across a continuous transition with increase in $T$.  From weak to moderately strong values of $U/t$, the transition temperature increases with $U/t$ in both the methods. But for $U>2\Delta$, while the transition temperature keeps increasing with $U$ within IPT, it goes as $t^2U/(U^2-4\Delta^2)$ within CTQMC, following the physics of the Heisenberg model as it should.

The rest of this paper is organized as follows. In Section $I$ we present the details of the model and the calculational methods used. Section $II$ describes in detail the $T=0$ phase diagram of IHM at half filling within IPT and CTQMC. Here we see a good qualitative and quantitative consistency between the two methods for a large range of parameters.
Section $III$ has details of the finite $T$ phase diagram within IPT and its comparison to that obtained using CTQMC. We focus specifically on the regime of extreme correlations, where CTQMC works well but IPT does not. We end this paper with conclusions and discussions. In appendices A and C we present detailed discussion on the nature of the phase transition involved while appendix B shows a comparison of results within the DMFT and the HF theory.

\section{Model and methods}
The model we consider has tight-binding electrons on a bipartite
lattice (sub-lattices A and B) described by the Hamiltonian
\[
H=-t\sum_{i\in A,j\in B,\sigma} [~c^{\dagger}_{i\sigma}c_{j\sigma}
+h.c~]+ \Delta \sum_{i\in A}n_{i} -\Delta\sum_{i \in B} n_{i} \]
\be
\mbox{~~~~~~~~~~~~~~~~~~~~~~~~~~~}+U\sum_{i}n_{i\ua}n_{i\da}-\mu
\sum_{i}n_{i} \label {model}
\ee
Here $t$ is the nearest neighbor
hopping, $U$ the Hubbard repulsion and $\Delta$ a one-body
staggered potential which doubles the unit cell. The chemical potential is
chosen to be $\mu=U/2$, so that the
average occupancy per site is $\left(\langle n_A
\rangle + \langle n_B \rangle \right)/2=1$
, corresponding to ``half-filling''.

\subsection{Dynamical Mean Field Theory (DMFT)}
 Here we study this model using the DMFT approach.
The DMFT approximation is exact in the limit of large
dimensionality \cite{georges,jarrell} and has been demonstrated to be successful in understanding the metal-insulator transition
\cite{georges,jarrell} in the usual Hubbard model, which is the
$\Delta =0$ limit of Eq.~(\ref{model}). We focus in this paper on
the anti-ferromagnetic sector of
Eq.~(\ref{model}), for which it is convenient to introduce the
matrix Green's function \be
\hat{G}_{\alpha\beta}^{\sigma}({\bf{k}},i\omega_n)
= \lbr \begin{array} {cc} \zeta_{A\sigma}({\bf{k}},i\omega_n) & -\epsilon_{\bf{k}} \\
-\epsilon_{\bf{k}} & \zeta_{B\sigma}({\bf{k}},i\omega_n)
\end{array}\rbr^{-1} \label{Greensfn} \ee where $\alpha,\beta$ are
sub-lattice ($A,B$) indices, $\sigma$ is the spin index, ${\bf k}$ belongs to the first Brillouin Zone (BZ) of \emph{one sub-lattice},
$i\omega_n=(2n+1)\pi T$ and $T$ is the temperature.
The kinetic energy is described by the dispersion
$\epsilon_{\bf{k}}$, and $\zeta_{A(B)\sigma} \equiv i\omega_n \mp
\Delta+\mu-\Sigma_{A(B)\sigma}(i\omega_n)$. Within the DMFT approach the
self energy is approximated as purely local~\cite{georges}. Thus the diagonal
self-energies $\Sigma_{\alpha\sigma}(i\omega_n)$ are ${\bf
k}$-independent and the off-diagonal self-energies vanish
(since the latter would couple the A and B sub-lattices).

The DMFT approach includes {\it local} quantum fluctuations by
mapping \cite{georges,jarrell} the lattice problem onto a
single-site or ``impurity'' with local interaction $U$
hybridizing with a self-consistently determined bath as follows.
(i) We start with a guess for $\Sigma_{\alpha\sigma}(\omega^{+})$, $m_s$ and
$\delta n$ and compute the local
$G_{\alpha\sigma}(i\omega_{n})=\sum_{{\bf{k}}}G_{\alpha\alpha}^{\sigma}({\bf{k}},i\omega_{n})$
rewritten as 
\be
G_{\alpha\sigma}(i\omega_n)=\zeta_{\bar{\alpha}\sigma}(i\omega_n)\int_{-\infty}^{\infty}
d\epsilon
\frac{\rho_{0}(\epsilon)}{\zeta_{A\sigma}(i\omega_n)\zeta_{B\sigma}(i\omega_n)-\epsilon^{2}}
\label{fullG} 
\ee 
where, for $\alpha=A(B)$,  $\bar{\alpha}=B(A)$, and $\rho_{0}(\epsilon)$ is the bare DOS for the
lattice considered (see below). We actually need to solve the
problem for only one sub lattice and use the relations
$G_{A\sigma}(i\omega_n) = -G_{B\sigma}(-i\omega_n)$ and $\Sigma_{A\sigma}(i\omega_n)
= U-\Sigma_{B\sigma}(-i\omega_n)$ valid at half filling. (ii) We next
determine the ``host Green's function"~\cite{georges,jarrell}
$\mathcal{G}_{0\alpha\sigma}$ from the Dyson equation
$\mathcal{G}_{0\alpha\sigma}^{-1}(i\omega_n) =
G_{\alpha\sigma}^{-1}(i\omega_n) + \Sigma_{\alpha\sigma}(i\omega_n)$. (iii)
We solve the impurity problem to obtain
$\Sigma_{\alpha\sigma}(i\omega_{n}) = \Sigma_{\alpha\sigma}
\left[\mathcal{G}_{0\alpha\sigma}(i\omega_{n})\right]$
(iv) We iterate steps (i), (ii) and (iii) till a self-consistent
solution is obtained.

\subsection{IPT as impurity solver}
 We use as our ``impurity solver'' in step (iii) a generalization
of the iterated perturbation theory (IPT) \cite{georges,kk} scheme
which has the merit of giving semi-analytical results directly in
the real frequency ($\omega^+ \equiv \omega + i0^+$) domain. The IPT
ansatz
$\Sigma_{\alpha\sigma}^{IPT}(\omega^{+})=\Sigma_{\alpha\sigma}^{HF}+A_{\alpha\sigma}\Sigma_{\alpha\sigma}^{(2)}(\omega^{+})$
is constructed to be (a) exact for $U/t \ll 1$, (b) exact for
$t/U=0$, and (c) exact in the large $\omega$ limit for all $U/t$,
which imposes various exact sum rules. Here $\Sigma_{\alpha\sigma}^{HF} = Un_{\alpha\bar{\sigma}}$
is the HF self energy with
\be 
n_{\alpha\sigma}=-\f{1}{\pi}\int_{-\infty}^{0}
d\omega~\mbox{Im~} G_{\alpha\sigma}(\omega^{+})~ , 
\ee
and 
\bea
\Sigma_{\alpha\sigma}^{(2)}(\omega^{+}) = U^{2}
\prod_{i=1}^{3}\int_{-\infty}^{\infty}d \epsilon_{i} ~~ [\tilde{\rho}_{\alpha\sigma}(\epsilon_{1}) \tilde{\rho}_{\alpha\bar{\sigma}}(-\epsilon_{2})\tilde{\rho}_{\alpha\sigma}(\epsilon_{3})]\nonumber\\
\times \f{\left[ f(\epsilon_{1})f(-\epsilon_{2})f(\epsilon_{3})
+f(-\epsilon_{1})f(\epsilon_{2})f(-\epsilon_{3}) \right]}{\omega^{+}-\epsilon_{1}+\epsilon_{2}-\epsilon_{3}}
.~ \label{sigma2}
\eea
This has the form of the second order self-energy with
$\tilde{\rho}_{\alpha\sigma}(\epsilon_{i})=-\mbox{Im}[\tilde{\mathcal{G}}_{0
\alpha\sigma}(\epsilon_{i}^{+})]/\pi$, where
$\tilde{\mathcal{G}}_{0\alpha\sigma}^{-1}(\omega^{+})=
\mathcal{G}_{0\alpha\sigma}^{-1}(\omega^{+})-\Sigma_{\alpha\sigma}^{HF}$ is
the Hartree corrected host Green's function and
$f(\epsilon)$ is the Fermi function. From condition (c) above we
find that $A_{\alpha\sigma}=
n_{\alpha\bar{\sigma}}(1-n_{\alpha\bar{\sigma}})/\left[n_{0\alpha\bar{\sigma}}(1-n_{0\alpha\bar{\sigma}})\right]$
with $n_{0\alpha\sigma} \equiv -\f{1}{\pi}\int
_{-\infty}^{0}~d
\omega~\mbox{Im}~\tilde{\mathcal{G}}_{0\alpha\sigma}(\omega^{+})~$. Note that at half filling, since $n_{A\sigma} = 1-n_{B\sigma}$, $A_{\alpha\sigma}$ is same for both the
sub lattices. 
For simplicity, here we present the results for the solution of the DMFT equations on a Bethe lattice of connectivity $z \rightarrow
\infty$. The hopping amplitude is re-scaled as $t\rightarrow t/\sqrt{z}$ to
get a non-trivial limit, and the bare DOS is then given by
$\rho_0(\epsilon) = \sqrt{4t^2-\epsilon^2}/(2\pi t^2),$ which
greatly simplifies the integral in Eq.~(\ref{fullG}).

\subsection{CTQMC as impurity solver}
\begin{figure}
\begin{center}
\includegraphics[width=3.0in,angle=0]{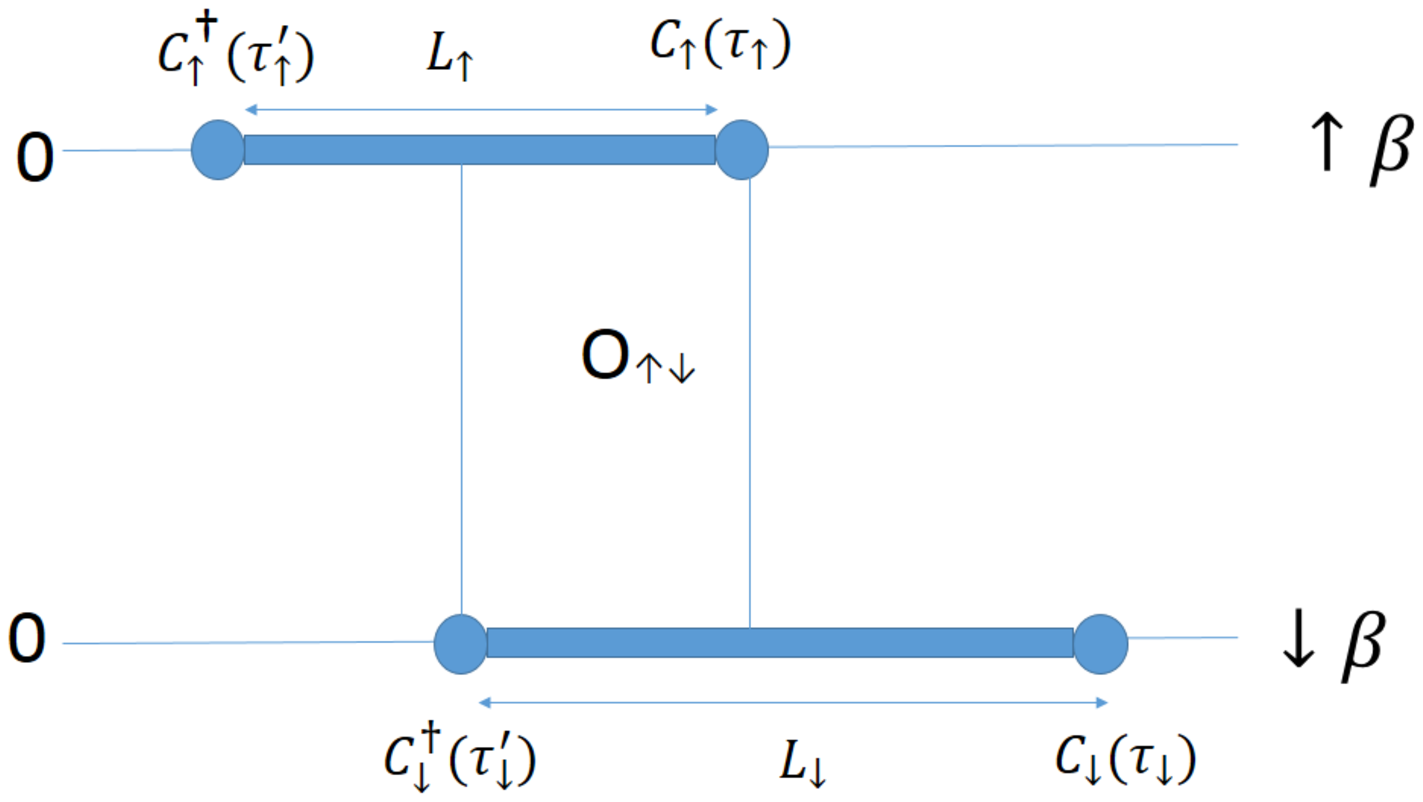}
\caption{A pictorial representation of a configuration generated by the CT-HYB algorithm with one up-spin and one down-spin electron. The total length of the segment in the $\tau$ space for which an electron with spin $\sigma$ lives is $L_\sigma$, and $O_{\uparrow\downarrow}$ is the total length of overlap (in $\tau$ space) for which electrons with both $\uparrow$ and $\downarrow$ spins are present.}
\label{ctqmc}
\end{center}
\vskip-6mm
\end{figure}

In this section we describe briefly the state-of-the-art impurity solver, the continuous time quantum Monte-Carlo (CTQMC) using the hybridisation expansion method~\cite{werner}, in the context of the IHM.   
The impurity model (IM) at site $\alpha$ corresponding to the IHM can be written as
\bea
H_{IM,\alpha} = \sum_{k\sigma}(\epsilon_k- s_\alpha\Delta)f^\dg_{k\alpha\sigma}f_{k\alpha\sigma}
+\sum_{k\sigma}V_{k\alpha}[f^\dg_{k\alpha\sigma}c_{\alpha\sigma} \nonumber \\
 +~h.c.]~+~Un_{\alpha\ua}n_{\alpha\da} -(\mu-s_\alpha\Delta)\sum_{\sigma}c^\dg_{\alpha\sigma}c_{\alpha\sigma}~~ 
\label{IM}
\eea
where $s_\alpha =1(-1)$ for $\alpha=A(B)$. $H_{IM,\alpha}$ describes the ``impurity'' in sub-lattice $\alpha$ coupled to the bath of $f$ fermions through the hybridisation term $V_{k\alpha}$.
It is straightforward to show, within a Grassmann functional integral formalism, that we can integrate out the fermionic bath variables in the partition function for the IM. After this step the partition function at site $\alpha$ becomes
\be
Z_\alpha=\int \mathcal{D} [c_{0\alpha\sigma} \d c_{0\alpha\sigma}]~~e^{-S_\alpha}
\label{zalpha}
\ee
where  $c_{0\alpha\sigma}\d$ and $c_{0\alpha\sigma}$ are Grassmann variables  representing the fermionic  ``impurity'' degrees of freedom at a site belonging to the $\alpha$ sub-lattice, and $S_\alpha$ is the functional,
\bea
S_\alpha=-\sum_{\sigma}\int_\bitato \mathrm{d}\tau \mathrm{d}\tau{\prime} c_{0\alpha \sigma} \d(\tau)\mathcal{G}^{-1}_{0\alpha \sigma}(\tau - \tau{\prime}) c_{0 \alpha\sigma}(\tau{\prime})\nonumber \\
+ \int_\bitato\mathrm{d} \tau U n_{\alpha\uparrow}(\tau) n_{\alpha\downarrow}(\tau)~~.
\eea
Here  $\mathcal{G}^{-1}_{0\alpha\sigma}(iw_n)$, the host Green's function at site $\alpha$, is related to the hybridisation amplitude $V_{k\alpha}$ via the relation

\be
\mathcal{G}^{-1}_{0\alpha \sigma}(iw_n)=iw_n + s_\alpha \Delta+\mu-\Delta_{\alpha\sigma}(i\omega_n)
\ee
where $\Delta_{\alpha\sigma}(i\omega_n) \equiv \sum_k \f{\vert V_k\vert^2}{i\omega_n-\epsilon_{k\sigma}+sign(\alpha)\Delta}$ is the hybridisation function. On the Bethe lattice of infinite connectivity, the self-consistent hybridization function for the IHM is given by $\Delta_{\alpha\sigma}(iw_n)=t^2G_{\bar{\alpha}\sigma}(iw_n)$, giving a simple relation between host Green's function and lattice Green's function as 
\be
\mathcal{G}^{-1}_{0\alpha \sigma}(iw_n)=iw_n + s_\alpha \Delta+\mu-t^2G_{\bar{\alpha}\sigma}(i\omega_n)
\label{G0_bethe}
\ee 
Hence $S_\alpha$ can be re-expressed as
\bea
S_{\alpha}=S_{loc}^\alpha+\sum_{\sigma}\int_0^{\beta}\mathrm{d}\tau  \mathrm{d}\tau{\prime}  c_{0 \alpha\sigma}\d(\tau)\Delta_{\alpha\sigma}(\tau - \tau{\prime}) c_{0\alpha \sigma}(\tau{\prime}) \nonumber \\
\equiv S_{loc}^\alpha + \sum_{\sigma} S^{\alpha\sigma}_{hyb}~~~~~~~~~~~~~~~~~~~~~~~~~~~~~~~~~~~~~~~~~~
\eea
where
\bea
S_{loc}^{\alpha}=\sum_{\sigma}\int_0^{\beta}\mathrm{d}\tau ~{c}_{0\alpha\sigma}\d{(\tau)}\left( {\frac{\partial }{\partial \tau}} -\mu + s_{\alpha} \Delta\right)\times \nonumber\\
{c}_{0\alpha\sigma}(\tau) + U\int_0^{\beta}\mathrm{d}\tau~ {n}_{0\alpha\uparrow}{(\tau)}{n}_{0\alpha\downarrow}(\tau)
\eea
The partition function, $Z_\alpha$  given by Eq.~{\ref{zalpha}} can then be expanded as a power series in $S^{\alpha\sigma}_{hyb}$  as,

\bea
Z_\alpha=Z_{0\alpha} \sum_{k} \frac{1}{k!^2}\int_0^{\beta}\mathrm{d}\tau_1 ...\mathrm{d}{\tau}_{k} \int_0^{\beta}\mathrm{d}{\tau^\prime}_1 ...\mathrm{d}{\tau^\prime}_{k} det \bf{\Delta_\alpha}
 \times \nonumber \\
{\langle T_{\tau} c_{0\alpha\sigma_1}(\tau_1) c_{0\alpha{\sigma^\prime}_1}\d({\tau^\prime}_1) ...c_{0\alpha\sigma_k}(\tau_k)c_{0\alpha{\sigma^\prime}_k} \d({\tau^\prime}_k)\rangle }_{S^{\alpha}_{loc}}~~
\eea
where

\be
\begin{split}
\bf{\Delta_\alpha}=
\begin{pmatrix}
  \Delta_{\alpha\sigma_1 {\alpha\sigma^\prime}_1}(\tau_1 , {\tau^\prime}_1) & ... & \Delta_{\alpha\sigma_1 {\alpha\sigma^\prime}_{k_\sigma}}({\tau_1} , {\tau^\prime}_{k}) \\
  ... & ... & ...\\
  ... & ... & ...\\
 \Delta_{\alpha\sigma_{k} {\alpha\sigma^\prime}_1}(\tau_{k} , {\tau^\prime}_1) & ... & \Delta_{\alpha\sigma_{k} {\alpha\sigma^\prime}_{k}}({\tau_{k}} , {\tau^\prime}_{k}) \\
 \end{pmatrix}
 \end{split}
\ee
and
\be
Z_{0\alpha} \equiv \int \mathcal{D} [c_{0\alpha\sigma} \d c_{0\alpha\sigma}] e^{-S^\alpha_{loc}}
\ee
In our case the matrix $\bf{\Delta}_\alpha$ is block-diagonal in ``up'' and ``down'' spin labels. Then the above equation simplifies to

\bea
\f{Z_\alpha}{Z_{0\alpha}}=\prod_{\sigma}\sum_{k_\sigma=0}^\infty
\frac{1}{k_\sigma!^2}\int_0^{\beta}\mathrm{d}\tau_{1}^\sigma ...\mathrm{d}\tau_{k_\sigma}^\sigma \int_0^{\beta}\mathrm{d}{\tau^\prime}_{1}^\sigma ...\mathrm{d}{\tau^\prime}_{k_\sigma}^\sigma  \nonumber \\
 det\bf{\Delta}_{\alpha\sigma} {\langle T_{\tau} c_{0\alpha\sigma}(\tau_{1}^\sigma)c_{0\alpha{\sigma}} \d(\tau_{1}^{\prime\sigma}) ...c_{0\alpha\sigma}(\tau_{k}^\sigma)c_{0\alpha{\sigma}} \d(\tau_{k_\sigma}^{\prime\sigma})\rangle }_{S^\alpha_{loc}} 
\eea
The CT-HYB algorithm generates ``configurations'' corresponding to the terms in Eq.~(16) with weights proportional to their contributions to the partition function Z. One such configuration is shown in Fig.~\ref{ctqmc}.\\

The CT-HYB algorithm can calculate important quantities such as the finite temperature imaginary-time Greens function, the density, the double occupancy etc. For example, the occupancy $n_{\alpha\sigma}$ is estimated from the average  length of all the segments: $n_{\alpha\sigma}$=$\langle L_{\alpha\sigma}\rangle_{MC}/\beta$; the double occupancy is obtained from the overlap $O_{\alpha\uparrow\downarrow}$ of segments as $D_\alpha=\langle O_{\alpha\uparrow\downarrow}\rangle_{MC}/\beta$; etc. For details  see~\cite{werner}.

The DMFT self consistency loop run as follows.
(1). One starts with a guess  for the local Green's function $G_{\bar{\alpha} \sigma}(iw_n)$ where for $\alpha=A,B$, $\bar{\alpha}=B,A$. (2). The host Green's function for the  $\alpha$ sub-lattice, $\mathcal{G}_{0\sigma\alpha}(iw_n)$, is calculated using Eq.~(\ref{G0_bethe}). (3). Using the host Green's function $\mathcal{G}_{0\sigma\alpha}(iw_n)$ the impurity solver calculates $G_{\alpha\sigma}(iw_n)$. Then step 2 is invoked again, and the process is repeated until $G_{A(B),\sigma}(iw_n)$ converges. We implement CT-HYB using TRIQS package~\cite{Triqs}.
\begin{figure}
\begin{center}
\includegraphics[width=3.0in,angle=0]{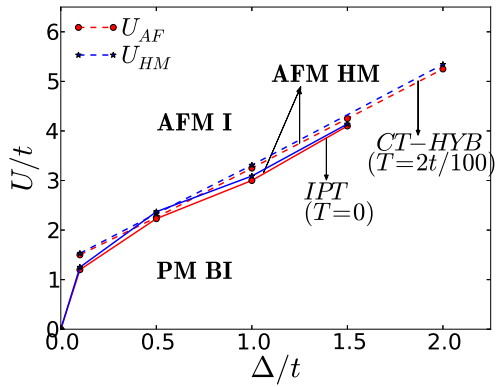}
\caption{Phase diagram of the model in Eq.~(\ref{model}) at half filling obtained using DMFT for the Bethe lattice with IPT at $T=0$ (solid lines) and CTQMC done at $T=0.02t$ (dashed lines). A
first order transition takes place at $U_{AF}$ such that for $U > U_{AF}$ the system has long range AFM
order, while for $U <U_{AF}$ it is a PM BI. For $U_{HM} > U_{AF}$, the spectral gap in one of the spin components
vanishes, resulting in a HM AFM phase for $U = U_{HM}$. For larger values of $U$ the system is an AFM insulator. Note that the transition points obtained using the two methods are in fairly good agreement with each other.} 
\label{phase_diag}
\end{center}
\vskip-6mm
\end{figure}
\section{$T=0$ phase diagram of the half-filled IHM}
The zero temperature phase diagram of the half-filled IHM obtained from the  DMFT+IPT study and the DMFT+CTQMC study (at $T=0.02t$) is shown in Fig.~\ref{phase_diag}. With increasing $U$ there occurs a first 
order transition between the PM BI and an AFM phase, characterized by a non-zero staggered magnetization $m_s$, at some threshold $U=U_{AF}$ (which is an increasing function of $\Delta$). Inside the AFM phase, a {\bf{half metal}} (HM) phase appears at $U=U_{HM} > U_{AF}$, where the gap in the single particle 
density of states (DOS) vanishes for one spin component while the other spin component has a non-zero spectral gap. When $U$ increases well above $U_{HM}$, the system becomes an AFM insulator (AFM I), where the gap in the DOS for both the spin components is controlled by, and increases linearly with, $U$. 

\begin{figure}[h!]
\begin{center}
\hspace{-0.5cm}
\includegraphics[width=2.75in,angle=0]{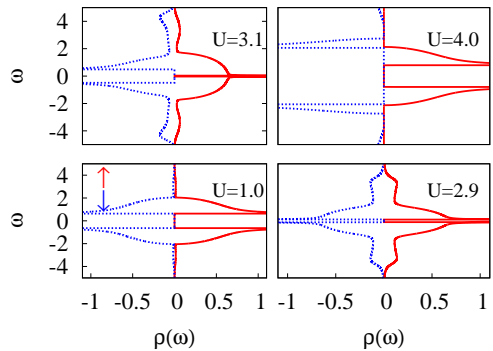}
\caption{The single particle DOS, $\rho_\sigma({\omega})$, calculated within DMFT+IPT, plotted as a function of $\omega$ for different values of $U/t$ for $\Delta =1.0t$ at $n=1$. The red curves are for the up spin component and  the blue dotted curves are for the down spin component. For $U<U_{AF} =3.0t$, the DOS is same for both the spin components, with a non zero spectral gap which decreases as $U/t$ increases, and the system is a PM BI. For $U > U_{AF}$, the DOS becomes different for the two spin components. At $U=3.1t > U_{AF}$, the DOS for the up-spin component has a vanishingly small gap  while the down spin DOS still has a finite gap. This is in close vicinity of the HM AFM point $U_{HM}$. At larger U values, there is a finite gap in the DOS for both the spin components,  and the system is an AFM I. }
\label{dos}
\end{center}
\vskip-6mm
\end{figure}
The phase diagram in Fig.~\ref{phase_diag} has been obtained from an analysis of various physical quantities, which we describe in detail below.

\begin{figure}[h!]
\begin{center}
\includegraphics[width=2.75in,angle=0]{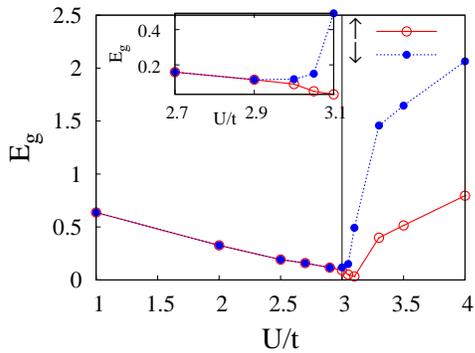}
\caption{The spectral gaps $E_{g\ua}$ and $E_{g\da}$, obtained from the DOS within DMFT+IPT, plotted as functions of U for $\Delta =1.0t$ at $n=1$. The red points are for the up-spin component and  the blue points are for the down-spin component. For $U<U_{AF}$ in the BI phase, $E_{g\ua}=E_{g\da}$ and both decrease with increasing $U/t$. At $U=U_{AF}$, there occurs a jump separating the two gaps, such that $E_{g\ua}$ is less than $E_{g\da}$. $E_{g\ua}$ becomes vanishingly small ($<0.01t$) at $U=3.1t$, close to the HM AFM point. Both 
$E_{g\ua}$ and $E_{g\da}$ increases with increase in $U/t$ in the AFM I phase ($U > U_{HM}$). The inset shows $E_g$ in the vicinity of the transition point.}
\label{Gap}
\end{center}
\end{figure}

{\bf{Single particle density of states (DOS):}}
In this subsection we discuss the single particle DOS $\rho_{\alpha,\sigma}(\omega) \equiv \ -\sum_k Im~\hat{G}_{\alpha\sigma}(k,\omega^{+})/\pi$, calculated using DMFT+IPT. Here $\alpha$ represents the sub lattice $A,B$ and $\sigma$ is the spin. Since at half filling $\rho_{A\sigma}(\omega)= \rho_{B\sigma}(-\omega)$, we will focus only on the total DOS $\rho_\sigma(\omega) = \rho_{A\sigma}(\omega)+\rho_{B\sigma}(\omega)$.  Fig.~\ref{dos} shows how $\rho_\sigma(\omega)$ evolves as a function of U for a fixed $\Delta=1.0t$. At small $U < U_{AF}$ (=3.0t for $\Delta=1.0t$), there is spin-symmetry in the DOS, and $\rho_\sigma(\omega)$ has a finite gap which decreases as $U/t$ increases, as shown in greater detail and clarity in Fig.~\ref{Gap}. We call this phase a PM BI as it is adiabatically connected to the $U=0$ band insulator. 

For $U > U_{AF}$, the spin symmetry in the DOS is lost as seen in the top  two plots of Fig.~\ref{dos}. 
The spectral gap in the up-spin component of the DOS is smaller than that for the down spin component, as can be seen more clearly in the inset of Fig.~\ref{Gap}. 
 We note that at half filling, even in the symmetry broken phase there is no net moment, i.e., $n_{\ua} = n_{\da}=1/2$. This is because of the symmetry relations of the Green's function (discussed earlier) which implies that $n_{A\sigma}=1-n_{B\sigma}$ and thus the total density of particles with spin $\sigma$ is $n_{\sigma} = \f{1}{2}[n_{A\sigma}+n_{B\sigma}] = 1/2$, although from the top two plots of Fig.~\ref{dos} it might seem that there is a net moment.  The point is that  Fig.~\ref{dos} shows only the {\it low $\omega$ DOS}, where the area under the DOS for the up-spin component is larger than that for the down spin component due to smaller spectral gap for the up-spin component. However, the weight loss for the down spin component in the low $\omega$ regime is compensated by its large $\omega$ part and the condition for no-net moment $n_{\sigma}=1/2$ holds. For $U > U_{AF}$, what the system has is a staggered moment, $m_s= n_{A\ua}-n_{A\da} = n_{B\da}-n_{B\ua}$, as discussed in more detail below.

 As $U$ increases above $U_{AF}$, the gap in the up-spin component of the DOS decreases rapidly, and becomes vanishingly small at a critical value $U=U_{HM}$ (equal to $3.09t$ when $\Delta=1.0t$), while the down-spin component still has a finite spectral gap (See Fig.~\ref{Gap}). Thus the IHM has a {\it half-metal} phase at a quantum critical point $U_{HM}$ embedded within the AFM regime. This prediction is further reinforced below from the low $\omega$ analysis of the spectral function. As $U$ increases further, the spectral gap in the DOS opens up again for the up-spin component as well, with both the spectral gaps increasing with $U/t$. This is the AFM insulating  (AFM I) phase. 
 
\begin{figure}[h!]
\begin{center}
%\vskip-1.0cm
\hspace{-0.5cm}
\includegraphics[width=2.5in,angle=0]{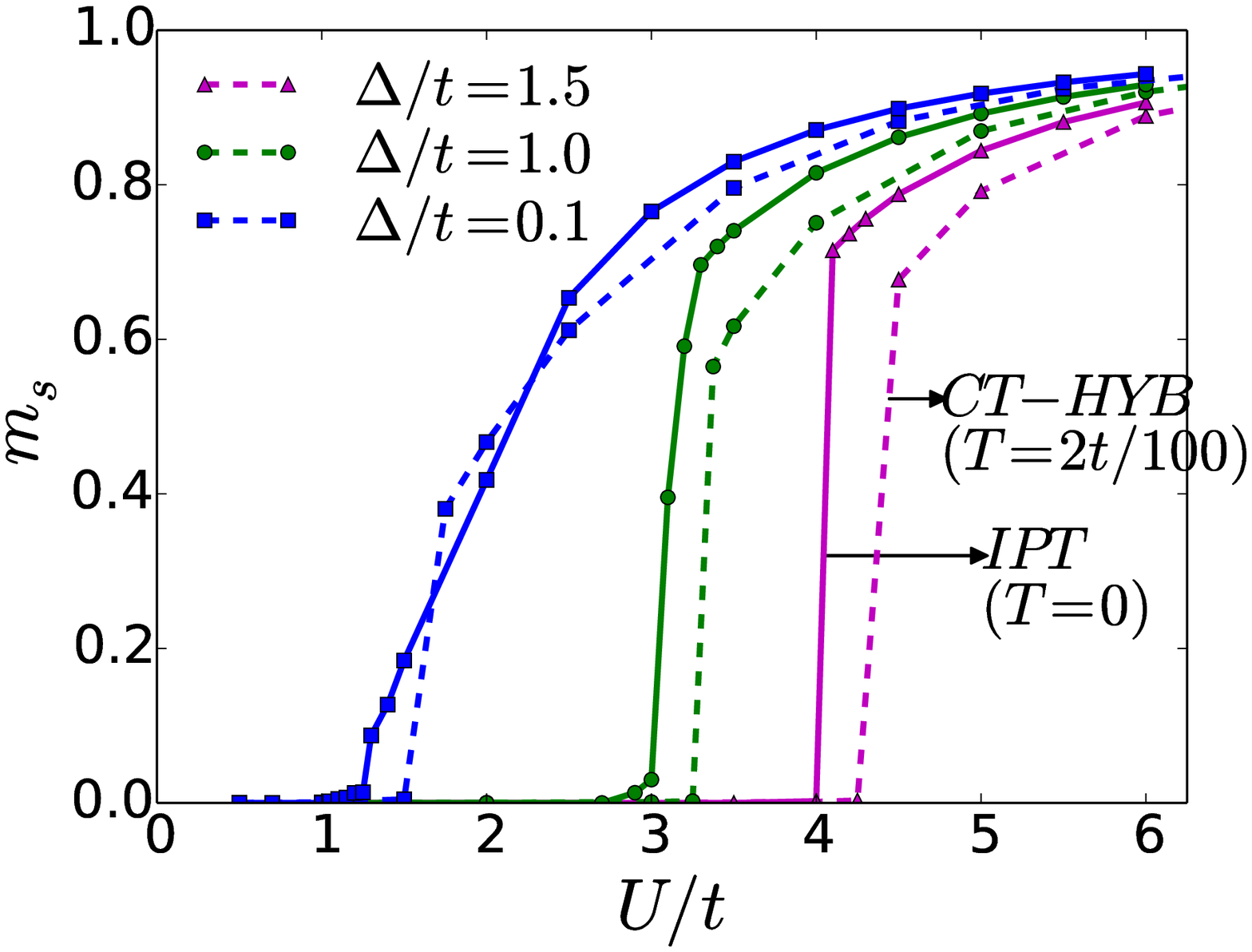}
\hspace{-2.25cm}
\includegraphics[width=2.5in,angle=0]{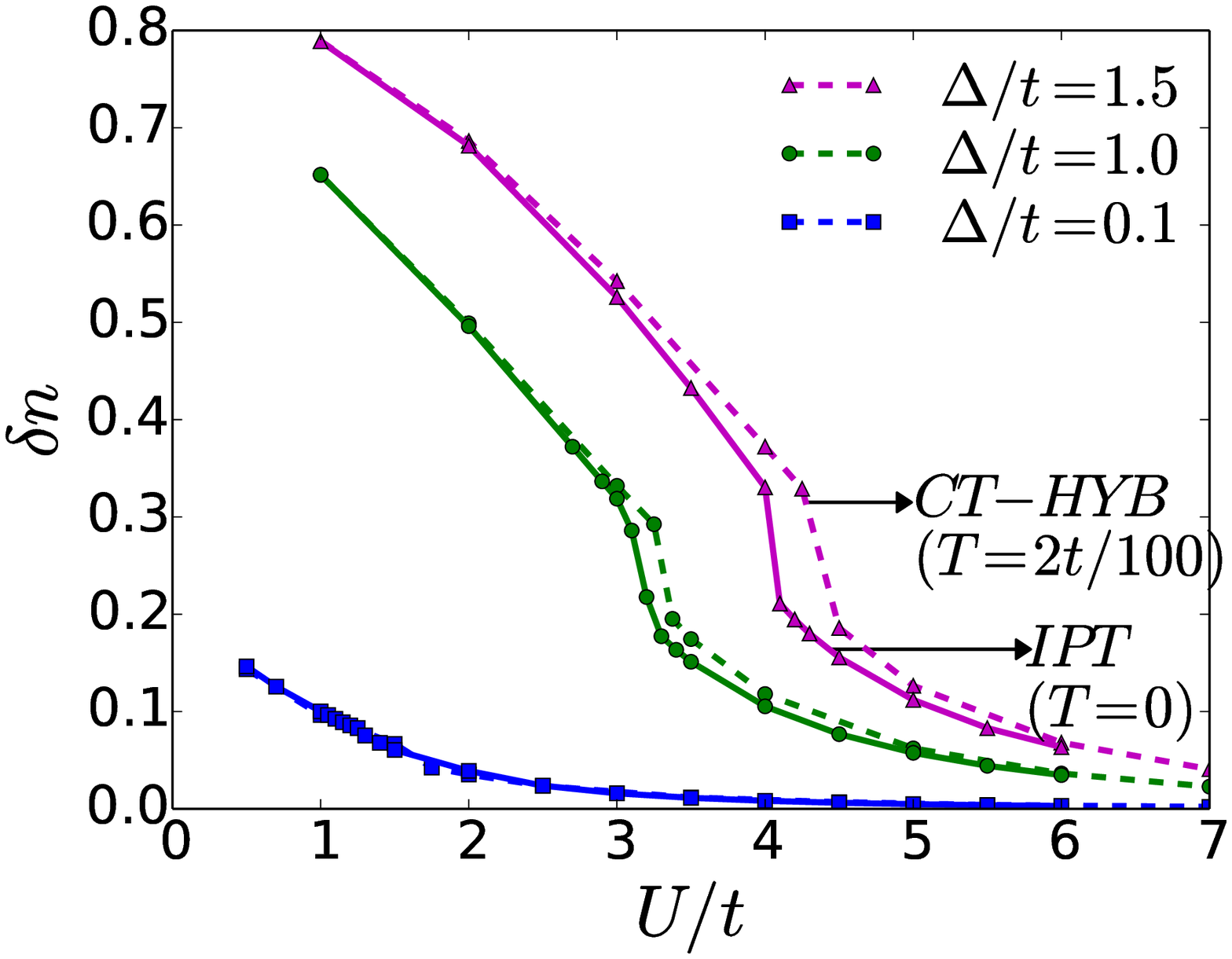}
\caption{Top panel: Staggered magnetization $m_s$ plotted as a function of $U/t$ at half-filling. A first order phase transition takes place with the onset of $m_s$ at $U_{AF}$.  Bottom Panel: Staggered occupancy $\delta n$ plotted as a function of $U/t$ at half-filling. $\delta n$ is non zero for all values of $U/t$ and a discontinuity occurs in $\delta n$ at $U_{AF}$. In both the panels the points connected with solid lines represent the data obtained from DMFT+IPT at $T=0$ and the points connected with dashed lines show data obtained within DMFT+CTQMC at $T=0.02t$. There is  quantitative consistency between the two methods for a range of $\Delta$ values. The phase transition is clearly first order in both the methods.}
\label{m_s}
\end{center}
\end{figure}

{\bf{Staggered magnetization and staggered occupancy:}}
The staggered magnetization $m_s$, defined as $m_s =(m_{zA}-m_{zB})/2$, calculated both within the DMFT+IPT ($T=0$) and DMFT+CTQMC (at $T=0.02t$) is shown in Fig.~\ref{m_s}. For a given value of $\Delta$, the staggered magnetization $m_s$ is zero below the corresponding $U_{AF}$ and becomes nonzero for larger $U$, with a discontinuous jump at $U_{AF}$ corresponding to a first order phase transition between the PM BI and the AFM phase.  Note that in the presence of the staggered potential, which opens up the gap in the DOS characteristic of the BI phase, the AFM instability does not occur unless  $U$ exceeds a finite thresh hold value $U_{AF}$.  The larger the value of $\Delta$, the larger is the value of $U$ required to overcome the effect of $\Delta$ and turn on the magnetisation. Thus both $U_{AF}$ and the jump in $m_s$ at $U_{AF}$ are increasing functions of $\Delta$.
The bottom panel of Fig.~\ref{m_s} shows the staggered occupancy, i.e., the difference in filling factor on the two sub lattices, defined as
$\delta n \equiv (n_{B}-n_{A})/2$.  Due to the staggered on site potential, this difference  is always non zero, even though the Hubbard $U$ tries to suppress it. For $U<U_{AF}$, $\delta n$ decreases monotonically and rapidly as
a function of $U$. At $U_{AF}$, there occurs a discontinuity in $\delta n$. For $U>U_{AF}$, $\delta n$ decreases more slowly with increasing U, but eventually becomes rather small in the AFM I phase, asymptotically approaching zero (as $t \Delta /U^2$ ) as $U \rightarrow \infty$.

Note that for all the $\Delta$ values, $m_s$ obtained using the CTQMC solver is slightly smaller than that from the IPT solver, while the transition point $U_{AF}$ obtained using CTQMC is larger than that within IPT. This is because CTQMC captures the effects of quantum fluctuations better than IPT. But overall, in the small to intermediate $U/t$ regime, there is  good quantitative correspondence between the low temperature CTQMC data and the $T=0$ data obtained within DMFT+IPT. Also, the nature of the phase transition is the same in both the methods.

 The results in Fig.~\ref{m_s} have been obtained  by  solving the DMFT+IPT equations starting from a small $U$ value and increasing $U$ slowly. When the DMFT equations are solved starting from a large $U$ guess and then decreasing $U$ slowly, one gets a different curve for $m_s$ (and also for $\delta n$)(Fig.~\ref{hysteresis}). A comparison of the ground state energies of these two spin-asymmetric solutions for the DMFT equation with the ground state energy of the PM sector shows that the real transition point $U_{AF}$ is the one where $m_s$ becomes non zero for the first time coming from the small $U$ side.  The hysteresis analysis discussed in Appendix A for $\Delta=1.0t$ confirms the nature of the transition from the PM to the AFM phase as being first order. But for very small values of $\Delta$, where both the transition point $U_{AF}$ and the jump in magnetisation at the transition point are very small, numerically it is difficult to see the nature of the transition. Since in the small $U$ regime, the Hartree-Fock (HF) theory also works well (as shown in Appendix B), we have carried out Ginzburg-Landau (GL) expansion of the ground state energy within the HF theory and confirmed that the phase transition from the PM to the AFM phase is of first order for any non zero $\Delta$ (for details see Appendix C). 

\begin{figure}[h!]
\begin{center}
\includegraphics[width=2.75in,angle=0]{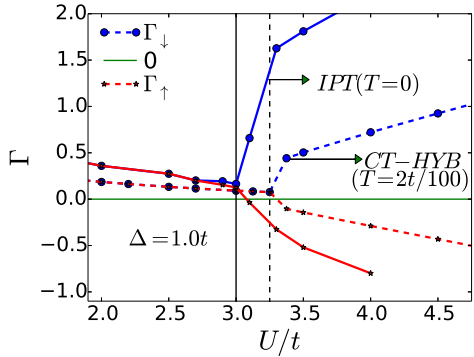}
\caption{$\Gamma_{\ua}$ and $\Gamma_{\da}$ (see Eq. 18) plotted as functions of $U/t$ for $\Delta=1.0t$. Points connected with solid lines are obtained within DMFT+IPT ($T=0$) and points connected with dashed lines are obtained within DMFT+CTQMC ($T=0.02t$). Note that $\Gamma_{\ua}$ changes sign within the AFM phase for $U>U_{AF}$ and crosses zero at $U_{HM}=3.09t$ within IPT and $U_{HM}=3.25t$ within CTQMC for $T=0.02t$. }
\label{Gamma}
\end{center}
\end{figure}
{\bf{Low $\omega$ analysis of the spectral function}}:
To understand the trend of the spectral gap and to confirm the existence of the HM AFM phase, we have carried out a low $\omega$ analysis of the self energy and the single particle spectral function.  The IPT self energy
$\Sigma_{\alpha\sigma}(\omega^{+}) \equiv \Sigma_{\alpha\sigma}^{\prime}(\omega)+i
\Sigma_{\alpha\sigma}^{\prime\prime}(\omega)$ has $\Sigma^{\prime\prime}_{\alpha}(\omega)$ vanishing for
$|\omega|\le 3E_{g\sigma}$ in both the insulating phases. This can
be understood from the imaginary part of $\Sigma_{\alpha\sigma}^{(2)}$ of
Eq.~(\ref{sigma2}), which comes from a three fermion final state. However, this is an artifact of ignoring collective modes (spin waves) within DMFT. In reality, since there are gapless spin waves that can be excited, the imaginary part of self energy will be zero only for $|\omega|\le E_{g\sigma}$ and the phase space constraints would make the result for $\Sigma^{\prime\prime}_{\alpha\sigma}$ just above threshold quite small. In the discussion below, we assume that  $\Sigma^{\prime\prime}_{\alpha\sigma}(\omega) = 0$ for $|\omega|\le E_{g\sigma}$.

In both the insulating phases, $\Sigma_{\alpha\sigma}^{\prime}(\omega)$
can be written at low $\omega$ as a Taylor expansion
$\Sigma_{\alpha\sigma}^{\prime}(\omega)=\Sigma_{\alpha\sigma}^{\prime}(0)+ \lbr 1-
Z_{\sigma}^{-1}\rbr \omega + \ldots$, where $Z_{\sigma}$ can be shown to be
independent of $\alpha$. The spectral function is defined by
$\mathcal{A}_{\alpha\alpha\sigma}(\epsilon,\omega)=\lbr -1/\pi\rbr {\rm Im}
G_{\alpha\alpha\sigma}(\epsilon,\omega^{+})$. Since $\Sigma_{\alpha\sigma}^{\prime\prime} = 0$ for $|\omega|\le 3E_{g\sigma}$, we find from Eq.~(\ref{Greensfn}) that
$A_{\alpha\alpha\sigma}(\epsilon,\omega) = \delta(r_{\sigma}(\omega)-\epsilon^2)$
with
$r_{\sigma}(\omega)=(\omega+\mu-\Delta-\Sigma^{\prime}_{A\sigma}(\omega))(\omega+\mu+\Delta-\Sigma^{\prime}_{B\sigma}(\omega))$.
As $\epsilon$ is real, $\omega$'s which satisfy $r_{\sigma}(\omega) < 0$ lie
within the gap. The energy gap is then given by $r_{\sigma}\left(E_{g\sigma}\right)= 0$ which, using the low-energy form of
$\Sigma_{\alpha\sigma}^{\prime}$ given above leads to the result 
\bea
E_{g\sigma} = Z_{\sigma}\vert \Delta -U/2+\Sigma^{\prime}_{A\sigma}(\omega=0)\vert \nonumber \\ 
=Z_{\sigma}\vert \Delta +U/2-\Sigma^{\prime}_{B\sigma}(\omega=0)\vert 
\label{gap}
\eea
where we have used the particle-hole symmetry.  
Let us write $\Sigma^{\prime}_{\alpha\sigma}(\omega=0) = S_{\alpha\sigma}+Un_{\alpha\bar{\sigma}}$ where the second term on the right hand side is the self energy within the Hartree-Fock approximation. 
Then one gets a more elaborate form for the expression of the gap, which is given below:
\bea
E_{g \sigma} = Z_{\sigma}\vert \Delta-{U/2 (\delta n +\sigma m_s)}+ S_{A,\sigma} \vert  \nonumber \\
= Z_{\sigma}\vert \Delta-{U/2 (\delta n +\sigma m_s)}-S_{B,\sigma} \vert \equiv Z_\sigma\vert \Gamma_\sigma \vert
\label{gap2} 
\eea

Fig.~\ref{Gamma} shows $\Gamma_{\sigma}$ as a function of $U$ for $\Delta=1.0t$ obtained within DMFT+IPT and DMFT+CTQMC($T=0.02t$). Within the CTQMC, the Green's function can be calculated only at Matsubara frequencies,  and thus the single particle DOS and the spectral gaps can not be obtained directly from the CTQMC data. But the low energy part of the self energy can be used to get an estimate of the spectral gap even from the CTQMC data. Specifically, in the CTQMC data, we have extrapolated the self energy to zero frequency and obtained approximate values of $S_{\alpha\sigma}$.
 
For $U<U_{AF}$, in the PM BI phase, $\Gamma_{\ua} = \Gamma_{\da} > 0$ and decreases as $U/t$ increases for a given $\Delta$.  At $U_{AF}$, $\Gamma_{\ua}$ becomes different from $\Gamma_{\da}$.  As $U/t$ increases further, within the AFM phase, $\Gamma_{\ua}$ decreases and becomes negative for  $U > 3.1t$ within IPT and $U>3.25t$ within CTQMC.   Thus it must pass through a zero, making $E_{g\ua}$ zero inside the AFM phase,  at $U_{HM}=3.09t(3.25t)$ for $\Delta=1.0t$ within IPT(CTQMC). 
On the other hand, $\Gamma_{\da}$ remains always positive, giving a non-zero spectral gap for the down-spin component for all values of $U/t$ including $U_{HM}$. Note that $Z_{\sigma}$ is always positive and less than one by definition, and that $\Gamma_\sigma$ obtained from CTQMC and IPT show good quantitative correspondence for $U < U_{AF}$. As $U$ increases further, $\vert\Gamma_\sigma\vert$ within IPT becomes much larger than that within CTQMC.
Within both the methods, we do see a novel, half metal AFM phase at $U_{HM} > U_{AF}$, inside the AFM phase of the correlated BI. Note that the half-metal AFM phase is missed out completely in a simple mean field theory like Hartree-Fock theory, though the BI to AFM transition is captured (see Appendix B).

\begin{figure}[h!]
\begin{center}
\includegraphics[width=2.75in,angle=0]{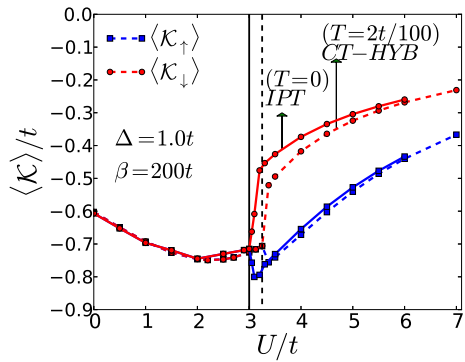}
\caption{ The kinetic energy $ \langle \mathcal{K}_{\ua}\rangle $ and $\langle \mathcal{K}_{\da}\rangle$ plotted as functions of $U/t$ for $\Delta=1.0t$. The points connected with solid line are obtained within DMFT+IPT ($T=0$) and those connected with dashed line are obtained within DMFT+CTQMC ($T=0.02t$). The kinetic energy decreases with increase in $U/t$ deep in the PM BI phase, which indicates states becoming more extended with increasing $U$, while it increases in the AFM I phase, suggesting localization.
For $U_{AF} < U < U_{HM}$, $\langle \hat{\mathcal{K}_\ua}\rangle$ decreases with $U/t$, reaching a minimum at $U_{HM}$, while $\langle \hat{\mathcal{K}_\da}\rangle$ increases with increase in $U/t$ as in the AFM I phase.}
\label{KE}
\end{center}
\end{figure}

{\bf{Kinetic energy}}: We have also studied the spin-resolved kinetic energy which is defined as $\langle \hat{\mathcal{K}_{\sigma}} \rangle = -\frac{2}{\pi}\int_{-\infty}^{0} d\omega~\int d\epsilon ~\epsilon ~\rho_0(\epsilon) Im G^\sigma_{AB}(\epsilon,\omega^{+})$. Fig.~\ref{KE} shows the $\langle \hat{\mathcal{K}}_{\sigma}\rangle$ obtained within DMFT+IPT ($T=0$) and DMFT+CTQMC ($T=0.02t$). To calculate the KE within DMFT+CTQMC, which gives the Green's function at fermionic Matsubara frequencies, we use $T\sum_n G^\sigma_{AB}(i\omega_n) = -\frac{1}{\pi}\int d\omega f(\omega) {\mathcal Im}G^\sigma_{AB}(\omega^+)$, where $f(\omega)$ is the Fermi distribution function, and derive the following version of the above expression for the KE:
\be
\langle \hat{\mathcal{K}_{\sigma}} \rangle = 2T~\int d\epsilon ~\epsilon ~\rho_0(\epsilon)\sum_{n} G^\sigma_{AB}(\epsilon,i\omega_n)
\label{ke_qmc}
\ee
where $G^\sigma_{AB}(\epsilon,i\omega_n)$ is the off-diagonal element of the full Green's function defined in Eq.~\ref{fullG}.

In the PM BI phase, as the spectral gap reduces with increase in $U$, $\langle \hat{\mathcal{K}_{\sigma}}\rangle$ decreases until the correlation starts pushing the spectral weight from low energy to higher energy region. Once this happens, even though the spectral gap is decreasing within the BI phase, there occurs a slight increase in $\langle \hat{\mathcal{K}_{\sigma}}\rangle$.

In the AFM I phase, the kinetic energy for both the spin components increases with increase in $U$ due to the increase in the spectral gap.
In the regime for $ U_{AF}<U<U_{HM}$, $\langle \hat{\mathcal{K}_{\ua}}\rangle$ decreases with increase in $U$ just like in small $U$ limit of the BI phase. On the other hand, $\langle \hat{\mathcal{K}_{\da}}\rangle$ starts increasing with $U$ like in the AFM I phase.
Note that the kinetic energy for the up-spin component is minimum at $U_{HM}$ where the spectral gap is zero for the up-spin component and we have a HM AFM.

\begin{figure}[h!]
\begin{center}
\hspace{-0.4cm}
\includegraphics[width=2.75in,angle=0]{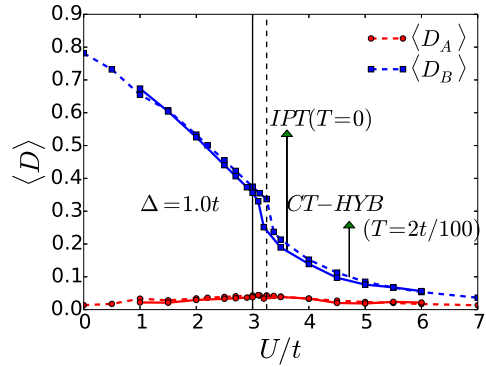}
\caption{Average double occupancy $D_\alpha = \la n_{\alpha\ua}n_{\alpha \da}\rangle$ on sublattice $\alpha=A,B$ vs $U$ for $\Delta=1.0t$. Points connected with solid line are calculated within DMFT+IPT($T=0$) and points connected by dashed line are calculated within DMFT+CTQMC ($T=0.02t$). Due to the staggered potential, $D_A \ll D_B$ for all values of $U/t$ with $D_B$ showing a monotonic decrease with $U/t$. Note that $D_\alpha$ within both the methods matches fairly well for $U \le 6t$.} 
\label{double_occ}
\end{center}
\end{figure}
{\bf{Double occupancy}}: The average double occupancy at site $\alpha$, $D_\alpha=\langle n_{\alpha\ua}n_{\alpha\da}\rangle$ can be calculated using the following equation within IPT:
\be
D_{\alpha} = \langle n_{\alpha\ua}n_{\alpha\da}\rangle=\f{1}{2U}[T \sum_{n,\sigma} i\omega_n G_{\alpha\sigma}(i\omega_n) + \mu_\alpha n_\alpha-\la \hat{\mathcal{K}}\ra]
\label{D}
\ee
 with $\alpha=B,A$ and $ \mu_\alpha \equiv (\mu + s_\alpha \Delta)$. Within CTQMC we calculated $D_\alpha$ by directly calculating the trace of $n_{\alpha\ua}n_{\alpha\da}$. Fig.~\ref{double_occ} shows $D_\alpha$ for $\Delta=1.0t$  at $T=0$ obtained within IPT ($T=0$) and CTQMC at $T=0.02t$. For the IHM, since a non zero $\Delta/t$ prefers to put more holes on the A sublattice and more double occupancies on the B sublattice, for all values of $U/t$, $D_A \ll D_B$. As $U/t$ increases $D_B$ shows a monotonic decrease with a discontinuity at $U_{AF}$. $D_A$ on the other hand, first increases slightly as $U$ increases below $U_{AF}$ and then starts decreasing with $U$. Fig.~\ref{double_occ} clearly shows that up to moderately strong values of $U/t$, the average double occupancy within IPT is quantitatively very close to that obtained within CTQMC. 

All of the above analysis shows clearly that for the $U/t$ range from weak to moderately strong, the IPT and CTQMC results match well. In our discussions in the following sections we focus on the differences between two approaches that arise when one looks at the extremely correlated regime of the IHM. 

\section{Extremely Correlated regime of the IHM and finite T phase diagram}
In this section we consider the extremely correlated regime of the IHM, namely, $U \gg t,\Delta$. 
In a regular Hubbard model ($\Delta=0$), the limit of $U \gg t$ effectively projects out doubly occupied sites from the Hilbert space. 
\begin{figure}[h!]
\begin{center}
\includegraphics[width=2.5in,angle=0]{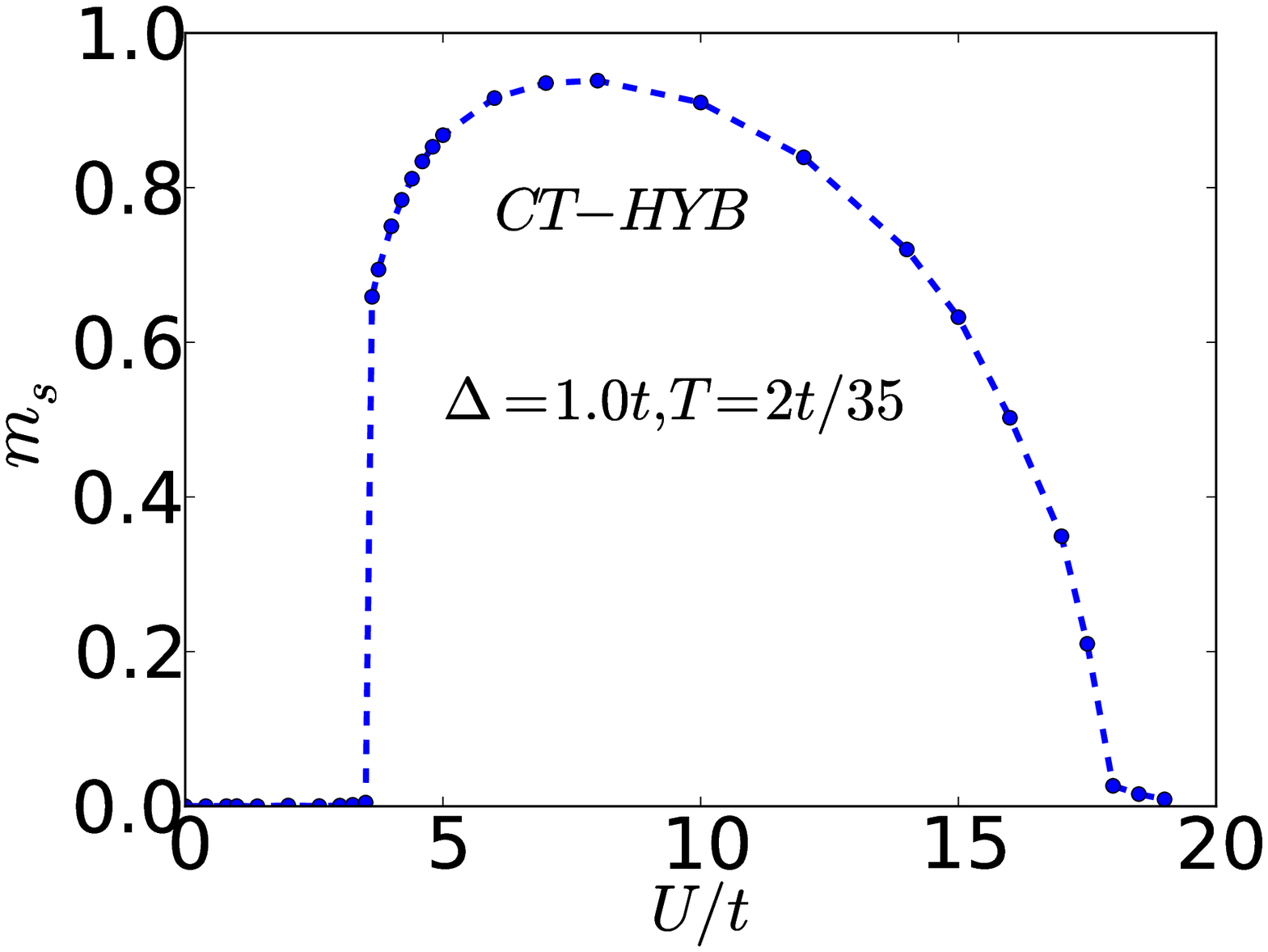}
\hspace{-0.2cm}
\includegraphics[width=2.25in,height=2.45in,angle=-90]{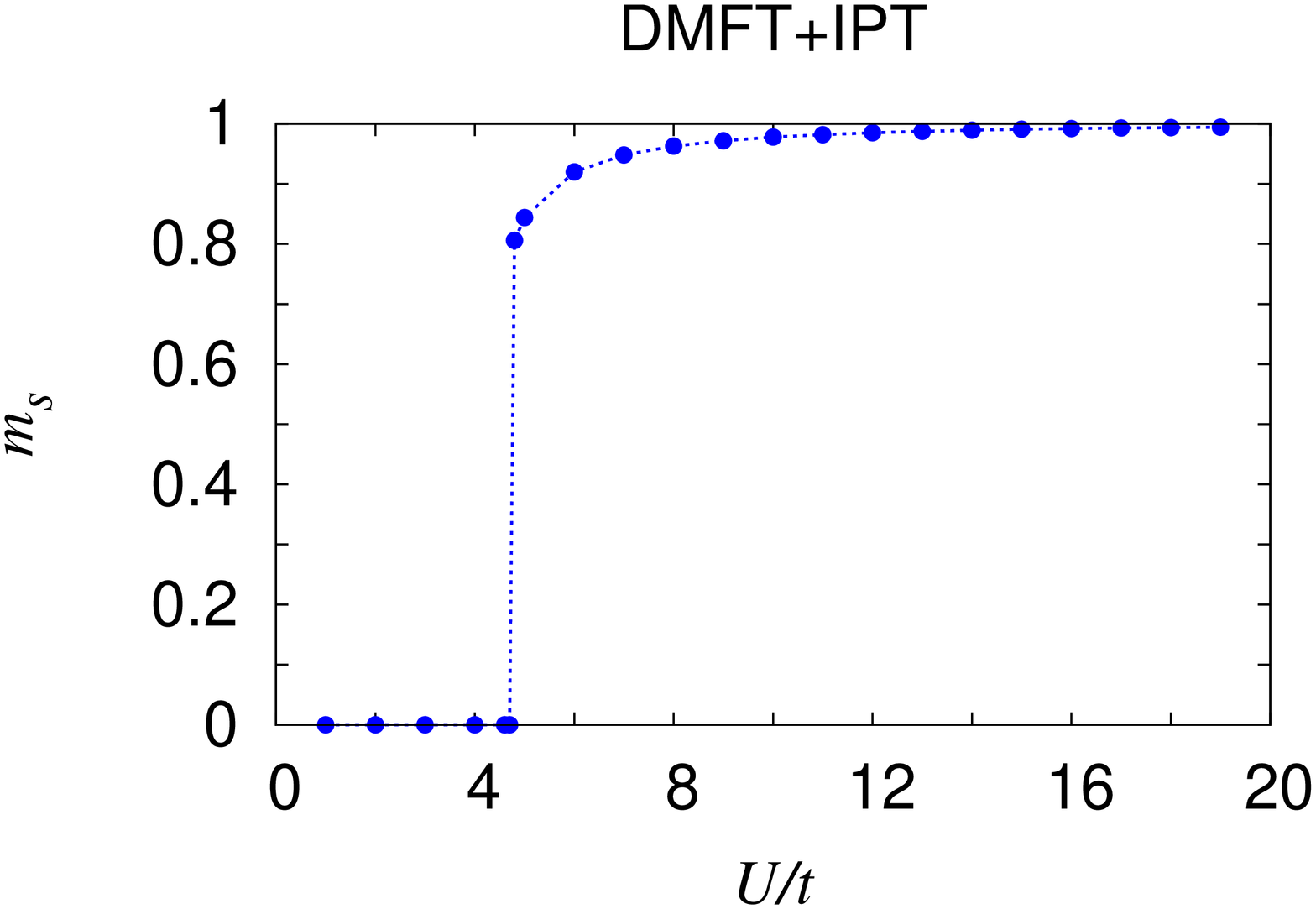}
\caption{Top: The staggered magnetisation vs $U/t$ for $T=2t/35$ and $\Delta=1.0t$ calculated within DMFT+CTQMC. As $U$ increases, first the AFM order turns on at $U_{AF}$ with a jump in $m_s$. As $U$ increases further, at $U_N \gg U_{AF}$, the AFM order goes to zero continuously. 
 Bottom: The staggered magnetisation $m_s$ vs $U/t$ for $T=0.4t$ and $\Delta=1.0t$ calculated within DMFT+IPT. As $U$ increases, $m_s$ keeps increasing and finally saturates to its maximum value.}
\label{qmcT}
\end{center} 
\end{figure}
For the IHM, at half filling, the energy cost for having a double occupancy on A(B) sublattice is $U \pm 2\Delta$. Thus, it is only for $U \gg t$ {\it{and}} $U \gg 2\Delta$ that one obtains the extremely correlated regime of the IHM where doubly occupied sites are projected out. 
In this limit, at half-filling, the effective low energy Hamiltonian for the IHM is again the Heisenberg model
\bea
H_{eff} = \tilde{J}\sum_{\la ij\ra} [S_i\cdot S_j-n_in_j/4],
\label{heff}
\eea
but the spin-exchange coupling is now $\tilde{J} = J/(1-x^2)$ with $x=2\Delta/U$ and $J=4t^2/U$. 
The Neel temperature $T_N$ of the Heisenberg model is proportional to $\tilde{J}$, which therefore depends upon $U$.  In dimensions higher than 2, for temperatures lower than $T_N$, the system has AFM ordering but the order is lost via a continuous transition as $T$ increases  past $T_N$.
In a finite $T$ calculation for the half-filled IHM, as $U$ increases beyond $2\Delta$, $\tilde{J}$ reduces and eventually at some $U_N$ where the corresponding $T_N(U=U_N) > T$, the magnetization is lost. Thus at any finite $T$, as one increases $U$, two phase transitions should be seen for the half-filled IHM. First at $U_{AF}$, where the magnetisation sets in via a first order transition, typically, and then at a larger $U_N > U_{AF}$ where the magnetisation is lost via a continuous phase transition.
\begin{figure}[h!]
\begin{center}
\hspace{-0.8cm}
\includegraphics[width=2.25in,height=2.5in,angle=-90]{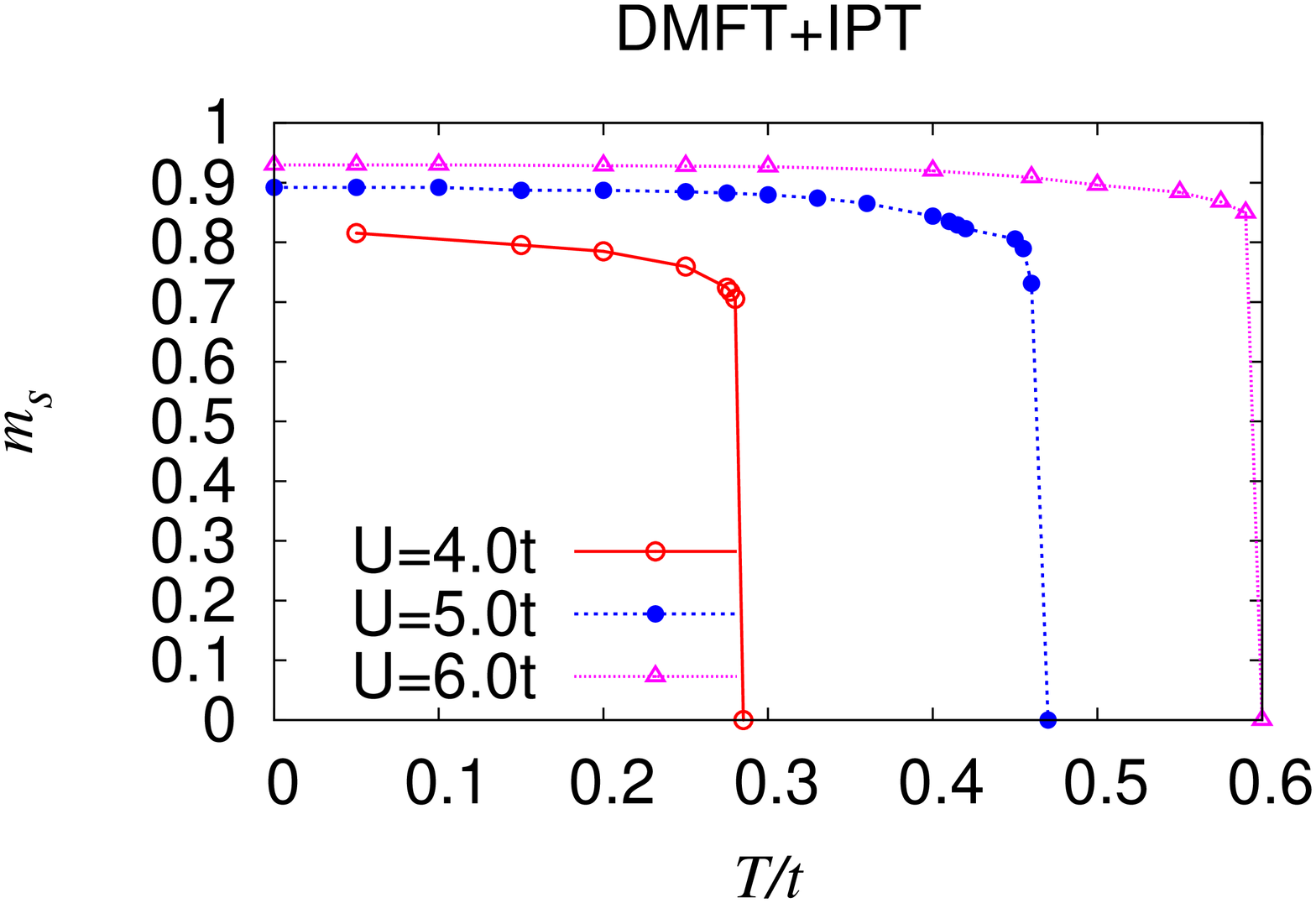}
\includegraphics[width=2.5in,angle=0]{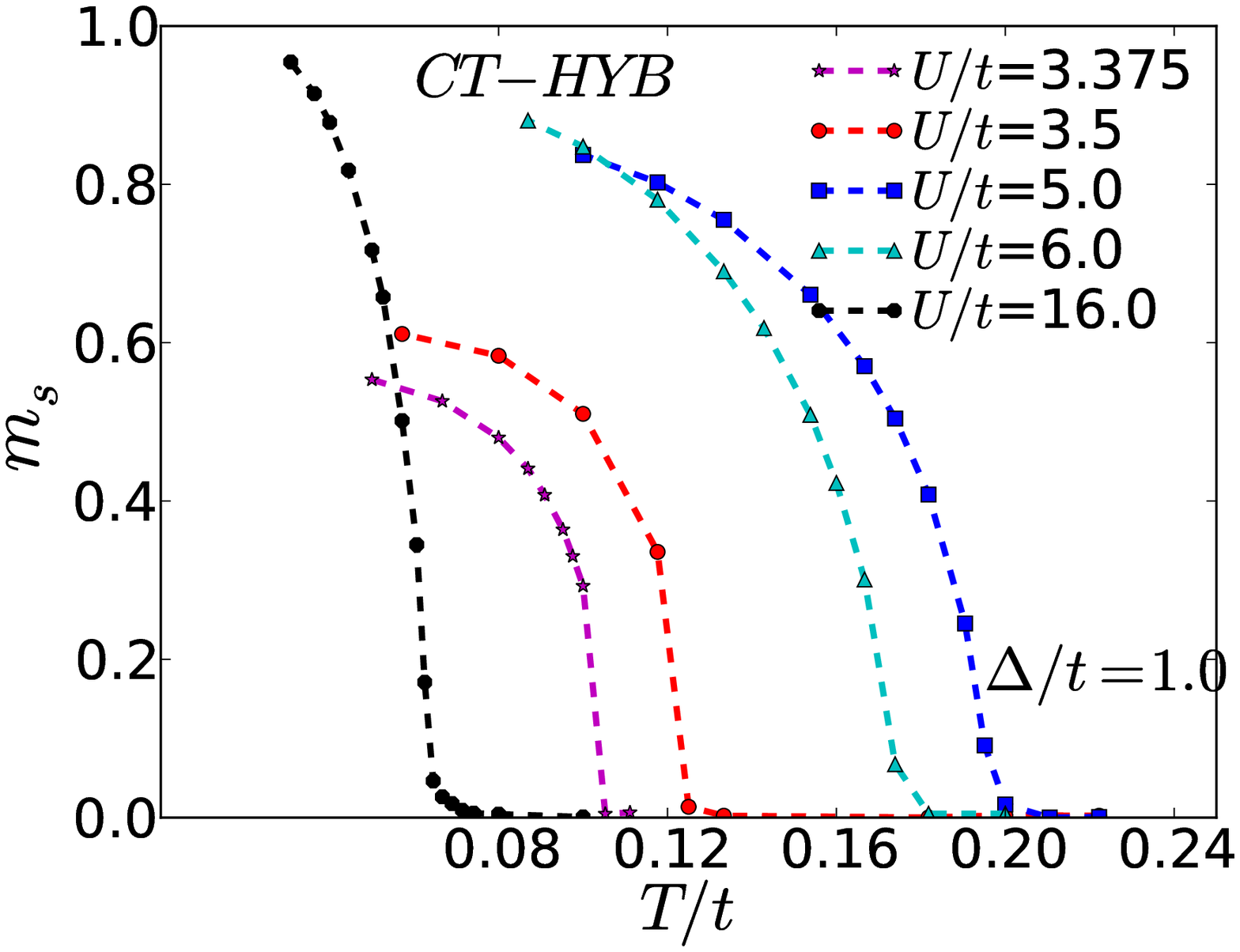}
\caption{Top: The staggered magnetization $m_s$ vs $T/t$ for $\Delta=1.0t$ and various values of $U$ obtained within DMFT+IPT. Bottom: $m_s$ vs $T/t$ obtained within CTQMC for $\Delta=1.0t$ for various values of $U$. Note that within IPT $m_s$ drops to zero via a first order phase transition at $T_{AF}$ which increases monotonically with $U/t$. But within CTQMC, for small values of $U$, though $m_s$ goes to zero via a first order transition but for larger values of $U/t$ there is a clear continuous transition as a function of $T$ in contrast to the IPT results. Also the transition temperature within CTQMC shows an increase with $U$ only upto $U=5.0t$ and starts decreasing with further increase in $U/t$.}
\label{finiteT}
\end{center}
\end{figure}

This is exactly what we see in the CTQMC result as seen in Fig.~\ref{qmcT}, which shows the staggered magnetization vs $U/t$ for $\Delta=1.0t$ and $T=2t/35$. We see that the staggered magnetisation turns on with a first order phase transition at $U_{AF}$, increases with increasing $U$ initially and then starts decreasing with further increase in $U$, finally vanishing at $U_N \gg U_{AF}$. But this second phase transition is not captured within DMFT+IPT, for which the finite $T$ the phase diagram is basically similar to the $T=0$ phase diagram.  Once the AFM order sets in at $U_{AF}$, as we keep increasing $U$ beyond $U_{AF}$, the staggered magnetization keeps increasing and never becomes zero, as shown in Fig.~\ref{qmcT}.  Thus though the suppression of double occupancy for large U is captured correctly to some extent within IPT, spin physics and the physics of the virtual hopping resulting in the effective Heisenberg model is not captured. Hence, while IPT interpolates between weak coupling to strong coupling regime (by satisfying the atomic limit), at the end it is a second order perturbation theory and especially for issues that crucially involve spin physics its validity breaks down in the regime of extremely strong correlations. 
\section{Thermal phase diagram}
Finally, we discuss how the AFM order is lost as the temperature $T/t$ increases for a fixed value of $\Delta$ and $U$. The top panel of Fig.~\ref{finiteT} shows the finite temperature results obtained within DMFT+IPT for $\Delta=1.0t$ and a few values of $U/t$. As shown here, the staggered magnetisation goes to zero via a clear first order phase transition at $T_{AF}$. On the other hand, as shown in the bottom panel of Fig.~\ref{finiteT}, within CTQMC the AFM order goes to zero via a first order transition, as the temperature $T/t$ increases, only for small values of $U/t$. For $U \gg 2\Delta$, the AFM order is lost continuously with a second order phase transition at $T_N$.

Further the transition temperature from the AFM phase to the PM phase has a very different dependence on $U$ and $\Delta$ within IPT and CTQMC, specially for $U \gg 2\Delta$. Within IPT, the transition temperature increases with increase in $U$ for a fixed $\Delta$ irrespective of whether we are in the intermediate coupling regime or in the regime of extreme correlations. To be more specific, it follows $U$, and does not follow $\tilde{J}$ for $U \gg t,\Delta$, whence the latter decreases with increase in $U$. This shows clearly that IPT does not capture the spin physics of extreme correlations correctly. 

\begin{figure}[h!]
\begin{center}
\hspace{-1.5cm}
\includegraphics[width=3.25in,angle=0]{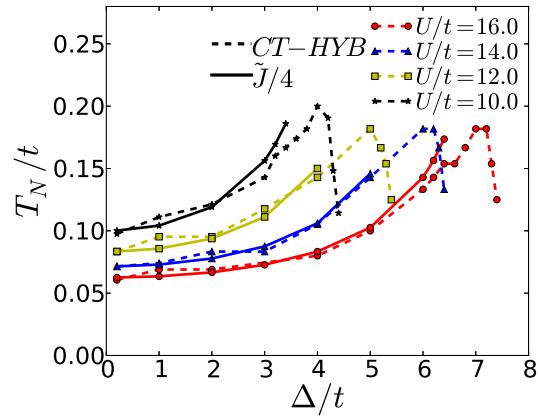}
\caption{Top: Neel temperature $T_N$ vs $\Delta/t$ obtained within DMFT+CTQMC for the IHM at half filling for various values of $U/t$. We see that for $\Delta \ll U$, $T_N$ follows $\tilde{J}/4$ very closely. But for $U \sim 2\Delta$, $T_N$ starts deviating from $\tilde{J}$ and  decreases with increase in $\Delta/U$.}
\label{NeelT}
\end{center}
\end{figure}

Within CTQMC, as is clear from Fig.~\ref{finiteT} for $\Delta=1.0t$, as $U/t$ increases, first the transition temperature $T_{AF}$ increases with increase in $U/t$ for $U/t < 5$. This trend is similar to what is seen within IPT. But as $U/t$ increases further, the physics of effective Heisenberg model starts playing a role and the transition temperature starts decreasing with further increase in $U$ as it is governed by $\tilde{J}$. 
For $U \gg 2\Delta$, as $\Delta$ increases, the spin-exchange coupling $\tilde{J}$ increases which is reflected clearly in the behaviour of $T_N$ in Fig.~\ref{NeelT}. These results are consistent with earlier DMFT+CTQMC work~\cite{rajdeep}. 
Fig.~\ref{NeelT} shows the transition temperature $T_N$ as a function of $\Delta/t$ for a few values of $U/t$. We have shown comparison of $T_N$ obtained within CTQMC with that of the Heisenberg model with spin exchange coupling of $\tilde{J}$. For $U \gg 2\Delta$, $\tilde{J}/4$ is a very good approximation to $T_N$. 
But for $U \sim 2\Delta$, $H_{eff}$ is not the correct low energy Hamiltonian of the model and we do not expect $T_N$ to be given by $\tilde{J}/4$. In fact in Fig.~\ref{NeelT}, we see that $T_N$ decreases as $\Delta$ increases beyond $U/2$ in contrast to what one would get from $\tilde{J}/4$.

\begin{figure}[h!]
\begin{center}
%\vskip-1.0cm
\hspace{-0.8cm}
\includegraphics[width=1.65in,angle=0]{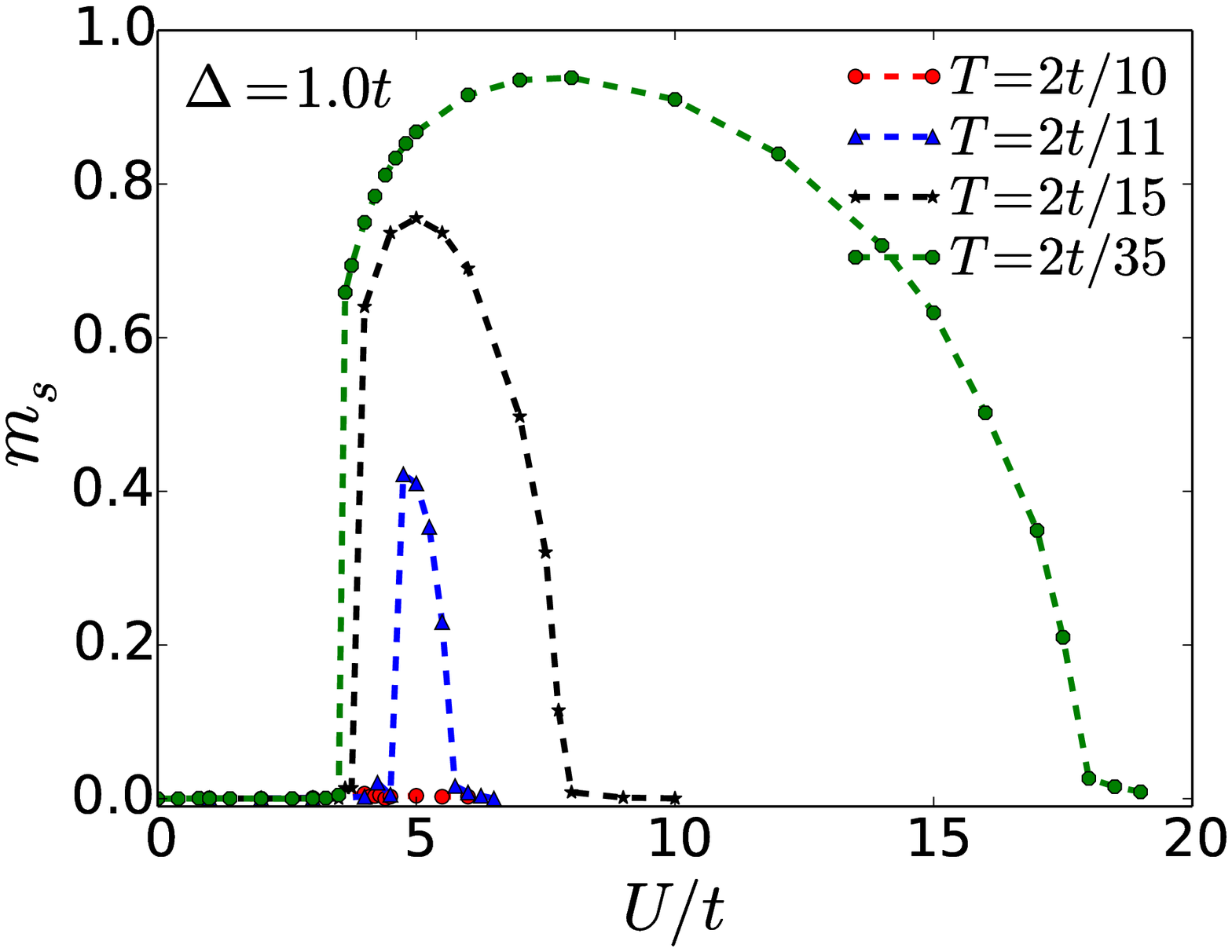}
\includegraphics[width=1.65in,angle=0]{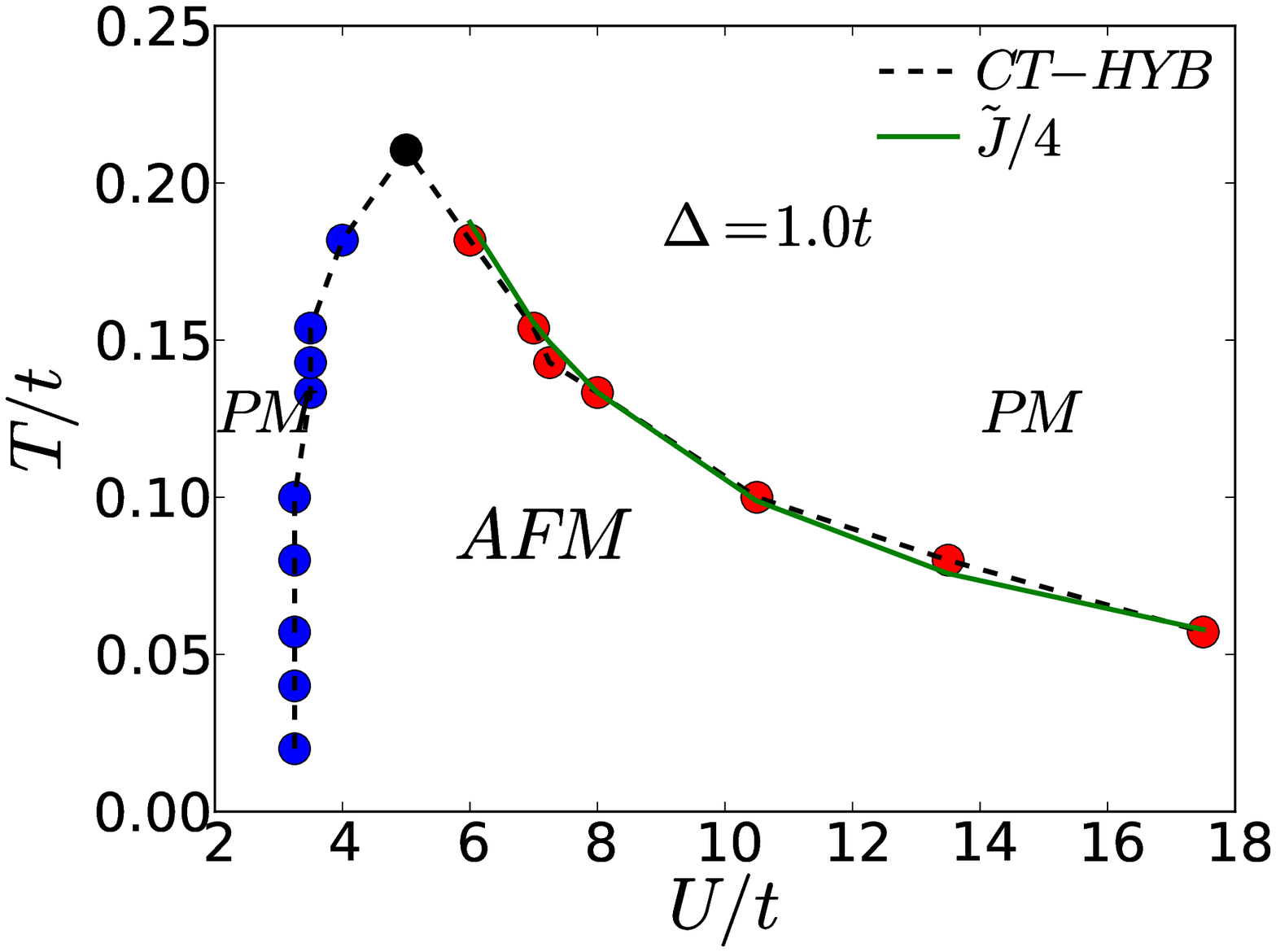}
\includegraphics[width=2.75in,angle=0]{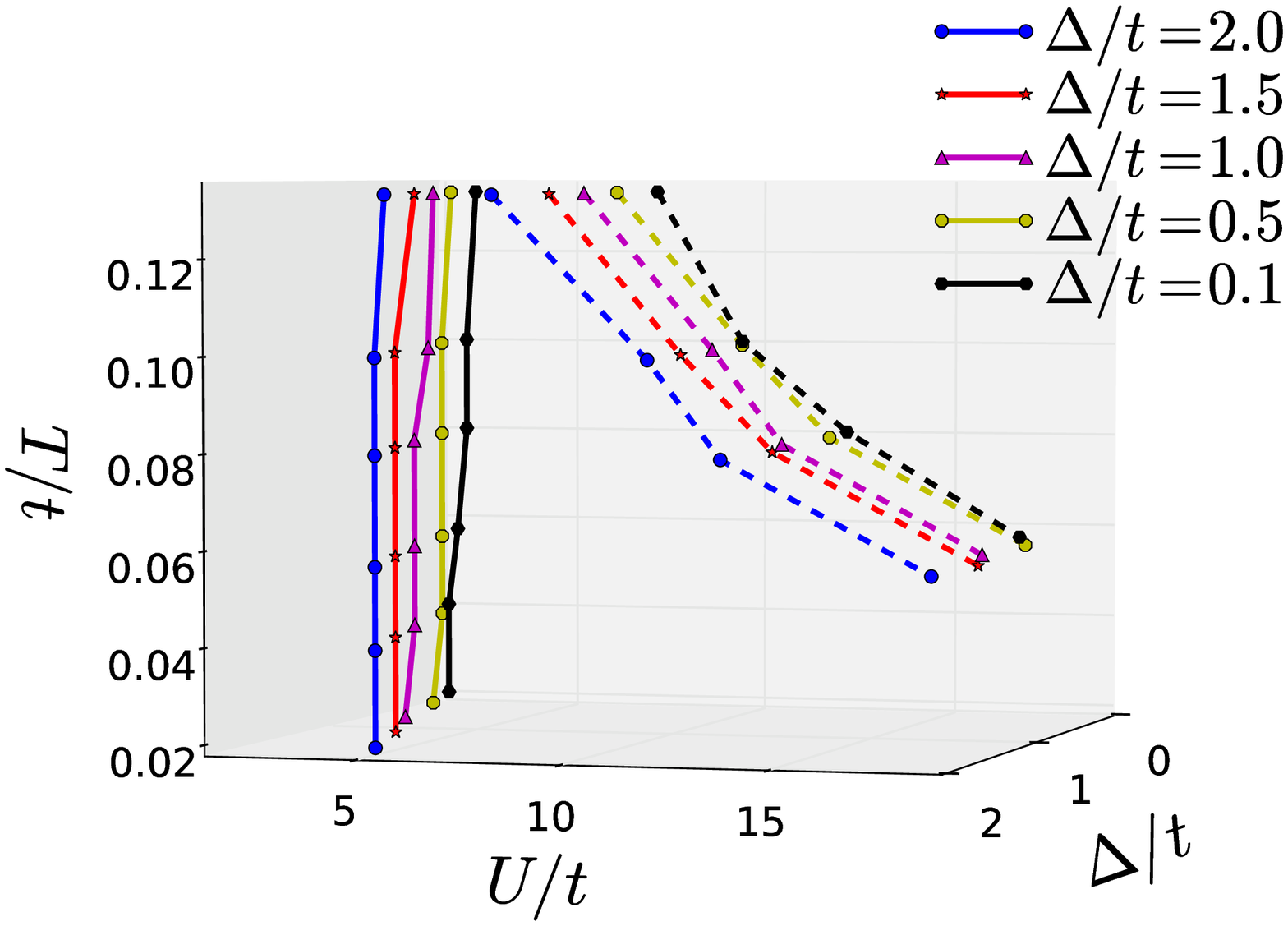}
\caption{Top: Left panel shows $m_S$ vs $U/t$ for $\Delta=1.0t$ and various values of $\beta$. These results are obtained within DMFT+CTQMC. The $m_S$ turns on via a first order transition at $U_{AF}(T)$ (shown as blue points in the right panel) while it is lost continuously at $U_N(T)$(shown as red points in the right panel). As $T$ increases, the $U$ range, $U_N(T)-U_{AF}(T)$, in which the system shows AFM order shrinks to zero. A tri-critical point, shown as a black point in the right panel, separates the lines of first and second order phase transitions. Bottom: Phase diagram for the  IHM at half filling in $T-U-\Delta$ space obtained within CTQMC. The surface made by the points connected by full lines is the first order transition surface from the PM to the AFM phase and the surface made by the points connected by dash lines is the second order transition surface from the AFM to a PM phase.}
\label{pd_qmc}
\end{center}
\end{figure}

Finally, we present the full magnetic phase diagram within DMFT+CTQMC inferred from calculations for a range of parameter values in Fig.~\ref{pd_qmc}. As shown in the bottom panel of Fig~\ref{pd_qmc}, in the 3 dimensional $T-U-\Delta$ space, there is a surface of first order phase transitions from PM BI to AFM insulator. Also there is a surface of second order phase transition across which the AFM order is lost continuously (although, as we have  noted, this surface does not show up in IPT). These two surfaces are separated by a line of tri-critical points. This can be seen more clearly in the top panel of Fig.~\ref{pd_qmc}. Here the left panel shows $m_s$ vs $U/t$ for various values of $T$. As $T$ increases, the value of $U_{AF}$ corresponding to the  first order transition, where the AFM turns on with a jump, increases. This is because there are more thermal fluctuations and a larger $U$ is required to stabilize the AFM order. Also, for the same reason, the AFM order does not survive for very small values of $\tilde{J}$ and thus the $U_N$ at which the AFM order is lost by a continuous transition decreases. These two transition points, namely $U_{AF}$ (point of first order phase transition) and $U_N$ (point of second order phase transition) come close as $T$ increases. There is a tricritical point which separates the two lines of first order and second order transitions. For $\Delta=1.0t$, from the CTQMC data we have generated, the tri-critical point seems to lie on the top of the dome of AFM region shown by a black point in the top-right panel of Fig.~\ref{pd_qmc}, but to be certain about this the calculations need to be done on a finer mesh of $U/t$ values.

\section{Discussion and Conclusion}
In conclusion, in this paper we have presented several new results from  a DMFT study of the ionic Hubbard model at half filling, i.e., the Hubbard model in the presence of a staggered potential, which makes the system a BI for $U=0$.  
As we turn on the on site repulsion $U$ in this BI, first an AFM order sets in via a first order transition at $U = U_{AF}$. This is followed by a quantum phase transition to novel half-metallic AFM phase at $U=U_{HM} > U_{AF}$. For still larger values of $U$, this system becomes an AFM insulator. 
Up to moderately strong values of $U$ (e.g., $U/t=6.0$ for $\Delta=1.0t$), the IPT captures the effects of electron-electron correlations quite well, and yields essentially the same results as CTQMC. But in the extremely correlated regime, where $U\gg \Delta,t$, DMFT+IPT does not work well, as becomes clear when one does a finite temperature study. 
At any finite $T$, while the IPT continues to show only one first order phase transition at which the AFM order turns on, the CTQMC shows, in addition, a second, continuous transition back to a PM phase, with its physics determined by the Heisenberg model.  As $T$ increases, the values of $U$ corresponding to the first and the second order transitions approach each other, shrinking the $U$ range  for which the long range AFM order is stable.  There is a line of tricritical point $T_{tcp}$ that separates the two surfaces of first and second order phase transitions. To the best of our knowledge, this feature of the IHM has not been discussed earlier in the literature. 

Recently there has been a DMFT+CTQMC study ~\cite{PM_ctqmc} of the half filled IHM within the PM sector, which shows a first order phase transition between Mott-Insulator and Metallic phases terminating at a critical point, just as in the Hubbard model at half filling. However, this critical point lies inside the dome of the AFM region shown in top right panel of Fig.~\ref{pd_qmc} and will be realised only if the AFM order is suppressed, either by lowered dimensionality (eg., quasi 2-d systems) or due to frustration (eg., by the presence of next nearest neighbour hopping, or a frustrated lattice). We hope to study these issues in future work. 
At the end we would like to mention that recently the IHM has been realised in ultracold fermions ~\cite{expt} on  a 2-dimensional honeycomb lattice and it can be extended to higher dimensional layered honeycomb lattice by introducing perpendicular hopping. Though our numerical study is on the Bethe lattice of infinite connectivity, we expect the qualitative physics to be the same for any bipartite lattice in $d \ge 2$ which has a compact density of states (DOS) like the DOS of the Bethe lattice of infinite connectivity. By choosing a large enough $\Delta$, it might be possible to realise an AFM phase for the IHM in experiments where the AFM order turns on with a first order transition and is lost by a second order transition by tuning $U$. It would be interesting to look for signatures of the various effects we have discussed, including the quantum phase transition, in the experimental measurements in such systems.  

\section{Acknowledgements}
We would like to thank the developers of TRIQS package which was used to carry out CTQMC using hybridisation expansion method used this work. S. B. would like to thank S.R.Hassan, V.B.Shenoy, Aabhaas Mallik and Prosenjit Haldar for many useful discussions. HRK acknowledges support by the DST, India, and the hospitality of the Department of Physics, UCSC, supported by the DOE under Grant No. FG02-06ER46319.
\section{Appendix A}
\begin{figure}[h!]
\begin{center}
\includegraphics[width=2.0in,angle=-90]{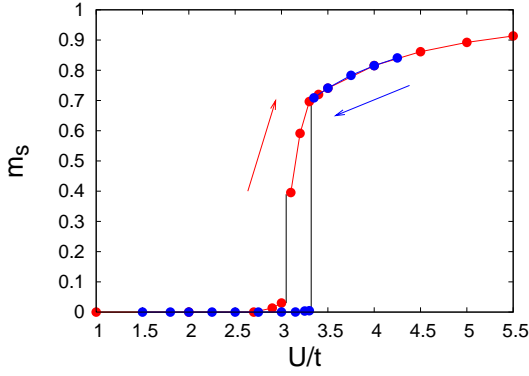}
\caption{Staggered magnetization $m_s$ plotted as a function of $U/t$ for $\Delta=1.0t$.
As  pointed by the arrows, the red curve is obtained by doing the DMFT+IPT calculation for increasing $U/t$ while the blue curve is obtained by starting from large $U/t$ side and decreasing $U/t$.}
\label{hysteresis}
\end{center}
\end{figure}
To characterize the nature of a phase transition, one normally studies its hysteresis behaviour.
We have solved the DMFT+IPT self-consistent equations, first starting from small $U$ value and increasing $U$ (AF-I) and then starting from large $U$ value and decreasing $U$ (AF-II). We see a clear hysteresis in the behavior of staggered magnetization $m_s$ and $\delta n$. In Fig.~\ref{hysteresis}, we have shown results for the staggered magnetization for $\Delta=1.0t$. We see that in the AF-I solution, $m_s$ becomes non-zero for $U > 3.0t$.  On the other hand, in AF-II solution, $m_s$ remains non-zero up to $U=3.2t$.
To get the transition point $U_{AF}$, we compare the ground state energy in the PM phase with that in the AF-I solution and AF-II solution. 
 \begin{figure}[h!]
\begin{center}
\includegraphics[width=3.2in,angle=0]{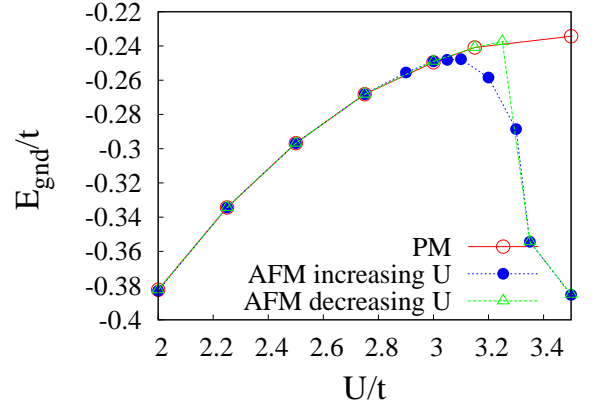}
\caption{ The ground state energy $E_{gnd}$ plotted as a function of $U/t$ for $\Delta=1.0t$. The red curve is $E_{gnd}$ obtained by solving the DMFT+IPT equation for model in Eq.~(\ref{model}) in the spin-symmetric sector while the blue and the green curves show the $E_{gnd}$ obtained by solving the DMFT+IPT equations in the AFM sector.  The blue curve is obtained by solving the self consistent equations coming from the small U side and the green one is obtained in decreasing U order. The AFM order sets in for $U> U_{AF}=3.0t$ for which the $E_{gnd}$ of the AFM sector is lower than that for the PM sector.}
\label{Egnd}
\end{center}
\end{figure}
Fig.~\ref{Egnd} shows the ground state energy for $\Delta=1.0t$ as a function of $U/t$. For $U<3.0t$, the PM phase is stable. For $U>3.0t$, $E_{gnd}$ for the AF-I sector  becomes lower than the ground state energy in the PM phase.  Notice that the $E_{gnd}$ of the AF-II sector becomes lower than the $E_{gnd}$ of the PM sector for larger value of $U/t$. Thus the AFM state becomes stable when for the first time $m_s$ becomes non-zero coming from the small $U$ side.  We call this point $U_{AF}$, which gives the boundary between PM BI and AFM phase in Fig.~\ref{phase_diag}.
\section{Appendix B}
For the model in Eq. [1] of the paper, the self energy within the Hartree-Fock (HF) approximation is given by
\bea
\Sigma_{A,\sigma} = U\la n_{A,\da}\ra = \f{U}{2}[1-\delta n+\sigma m_s] \nonumber \\
\Sigma_{B,\sigma} = U\la n_{B\da}\ra = \f{U}{2}[1+\delta n-\sigma m_s]
\eea

Here $m_s=(m_{zA}-m_{zB})/2$ is the staggered magnetisation with $m_{z\alpha}= n_{\alpha\ua}-n_{\alpha\da}$ and $\alpha=A,B$ is the sublattice index. 
$\delta n =(n_B-n_A)/2$ is the staggered occupancy, i.e., the difference in the filling factor of the two sublattices.
\begin{figure}
\begin{center}
\includegraphics[width=2.85in,angle=0]{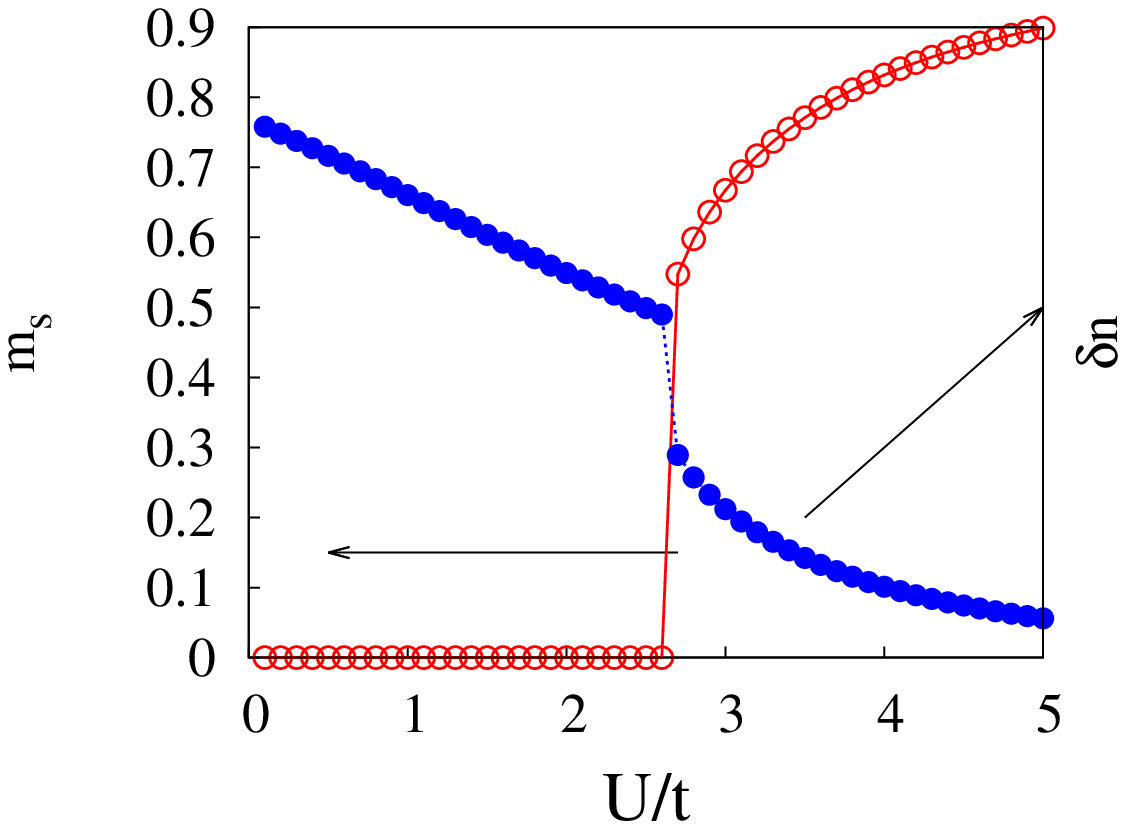}
\includegraphics[width=2.85in,angle=0]{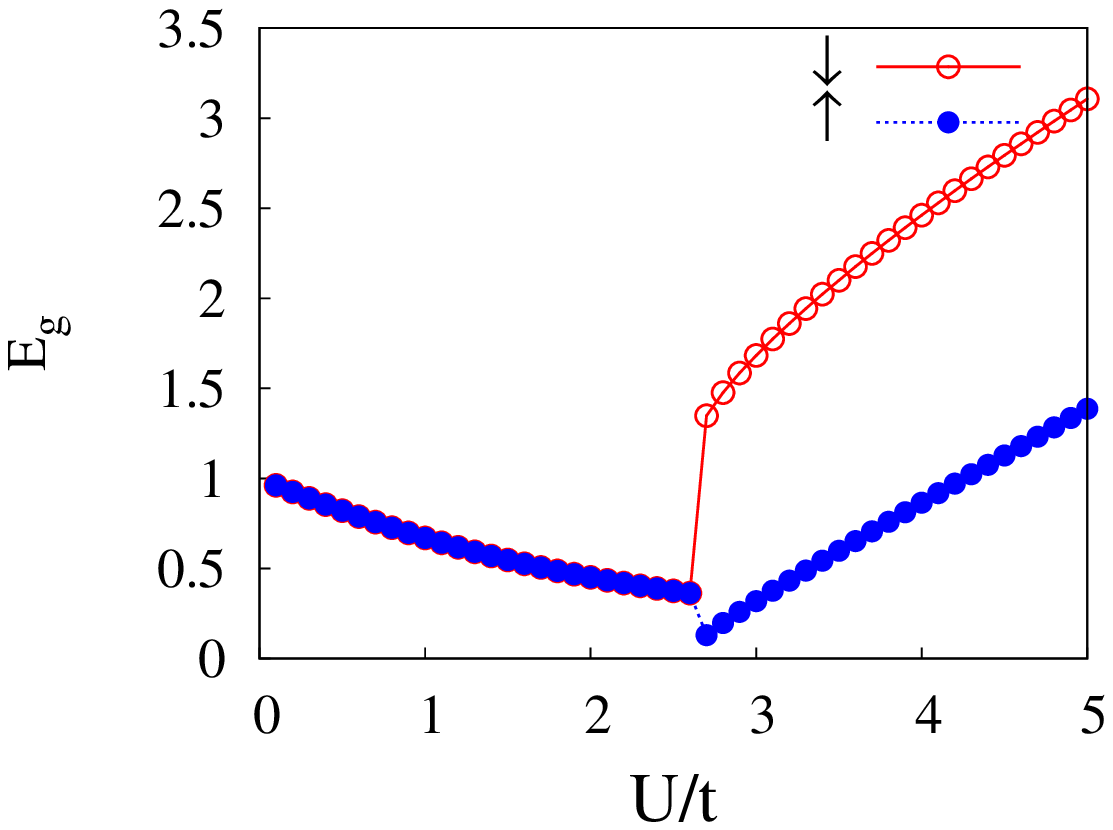}
\caption{Top: Plots of $m_s$ and $\delta n$ vs $U/t$ for $\Delta=1.0t$ and $n=1$ within the HF theory. Bottom: Spectral gap $E_{g\sigma}$ for the up and down spin components within the HF theory. As soon as the magnetic order turns on, both $E_{g\ua}$ and $E_{g\da}$ start increasing with $U/t$. Thus the half-metal phase, seen in the DMFT calculation just after the onset of the AFM order, is missing here and the system is an AFM insulator for all $U > U_{AF}$.}
\label{hf1}
\end{center}
\end{figure}
Since the bare Green's function (with U=0) is given by,
\be
\hat{G_0}_{\sigma}({\bf{k}},i\omega_n)
= \lbr \begin{array} {cc} i\omega_n+\Delta+\mu & -\epsilon_{\bf{k}} \\
-\epsilon_{\bf{k}} & i\omega_n-\Delta+\mu
\end{array}\rbr^{-1},  \label{G0} \ee
the HF corrected Green's function is given by
\be
\hat{G}_{\sigma}({\bf{k}},i\omega_n)
= \lbr \begin{array} {cc} i\omega_n+g_\sigma+\tilde{\mu} & -\epsilon_{\bf{k}} \\
-\epsilon_{\bf{k}} & i\omega_n-g_\sigma+\tilde{\mu}
\end{array}\rbr^{-1} \label{G} 
\ee
Here $\tilde{\mu}= \mu-\frac{U}{2}=0$ is the chemical potential and $g_{\sigma} =  \Delta-\f{U}{2}(\delta n+\sigma m_s)$  which gives a gap $E_{g\sigma}=|g_\sigma|$ in the single particle spectrum of $\sigma$ spin component. 
Using this Green's function, one gets the following self consistent equations for the physical quantities defined above:
\bea
m_s= \f{1}{2}\int d\epsilon \rho_0(\epsilon)\sum_\sigma \f{\sigma g_\sigma}{E_{\sigma}(\epsilon)}[f(E_\sigma(\epsilon))-f(-E_\sigma(\epsilon))] \\
\delta n= \f{1}{2}\int d\epsilon \rho_0(\epsilon) \sum_\sigma  \f{g_\sigma}{E_\sigma(\epsilon)}[f(-E_\sigma(\epsilon))-f(E_\sigma(\epsilon))] \\
n = \f{1}{2}\int d\epsilon \rho_0(\epsilon) \sum_\sigma  [f(E_\sigma(\epsilon))+f(-E_\sigma(\epsilon))] 
\eea
Here $E_\sigma(\epsilon) = \sqrt{\epsilon^2+g_\sigma^2}$, $f(E_\sigma(\epsilon)) = \f{1}{\exp(\beta(E_\sigma(\epsilon)-\tilde{\mu}))+1}$ is the Fermi function and $\rho_0(\epsilon)$ is the bare density of states of the lattice under consideration.

We have solved the self-consistent equations for the Bethe lattice of infinite connectivity and the results obtained at half-filling ($n=1$) and zero temperature are as follows.
\begin{figure}
\begin{center}
\includegraphics[width=3.5in,angle=0]{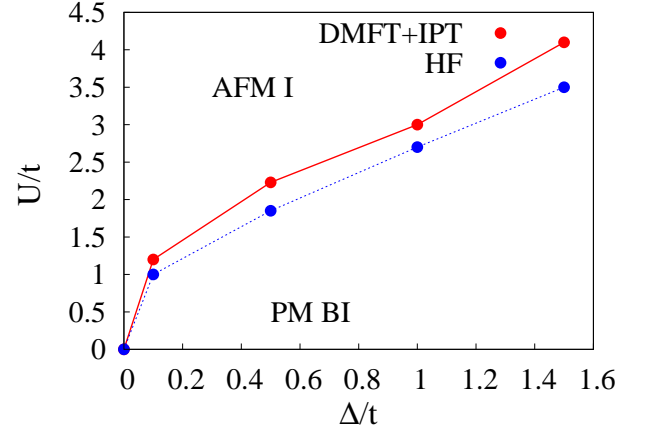}
\caption{Phase diagram at half filling at $T=0$ for Bethe lattice of infinite connectivity. Red circles are the data obtained from DMFT+IPT study while the blue circles are the data obtained from the HF theory. }
\label{hf2}
\end{center}
\end{figure}
For small $U/t$ the system is a BI with $m_s=0$ and  a non zero $\delta n$. At $U=U_{AF}$ a first order phase transition takes place with a jump in $m_s$ to a non zero value as shown in Fig.~\ref{hf1}. For $U>U_{AF}$, the system is an AFM insulator.
Fig.~\ref{hf2} shows the phase diagram at half-filling within the HF theory. For comparison we have also shown the phase diagram within DMFT+IPT at half-filling. The threshold $U_{AF}$ required to turn on the magnetisation is smaller in the HF theory as compared to its value within the DMFT+IPT. This is because quantum fluctuations captured in DMFT are missing in the HF theory; as an effect the magnetic order survives up to smaller values of $U$.  
However, the HF phase transition line approaches the DMFT line as $\Delta/t$ gets smaller. The bottom panel of Fig.~\ref{hf1} shows the spectral gaps $E_{g\sigma}$. Just after the AFM order sets in, the spectral gaps for both the spin components start increasing with $U/t$ which is in contrast to what is seen in the DMFT calculation, where the gap for one of the spin component keeps decreasing with $U/t$ even for $U> U_{AF}$ leading to a HM point at $U_{HM}> U_{AF}$. Thus within the simple HF theory, where the self energy is independent of $\omega$, there is no half-metallic phase at half filling. 
\section{Appendix C}
The ground state energy within the HF theory is
\be
E_{gnd} = -\sum_k E_\ua(\epsilon_k) -\sum_k E_\da(\epsilon_k)-U\sum_{\alpha=A,B} \la n_{\alpha \ua}\ra \la n_{\alpha \da}\ra 
\ee
where, as before, $E_{\sigma}(\epsilon_k)=\sqrt{g_\sigma^2+\epsilon_k^2}$ with $g_\sigma = \Delta-\frac{U}{2}(\delta n+\sigma m_s)$.
The last term in $E_{gnd}$ can be re-expressed as $\frac{U}{2}(\delta n^2-m_s^2)$.
Following the Ginzburg-Landau (GL) theory, we do the Taylor series expansion of $E_{gnd}$ for small $m_s$:
\bea
E_{gnd} \sim E_0 +m_s^2 E^{\prime\prime}(m_s=0)+m_s^4 E^{\prime\prime\prime\prime}(m_s=0) + ....\nonumber \\
= E_0 +\f{a}{2} m_s^2 + \f{b}{4} m_s^4 + \f{c}{6}m_s^6 +...
\label{GLE}
\eea
\begin{figure}
\begin{center}
\includegraphics[width=2.0in,angle=0]{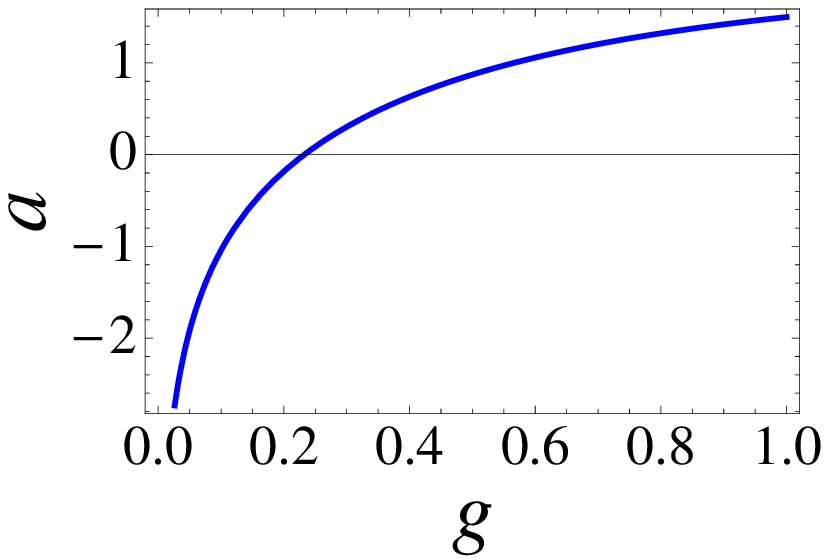}
\includegraphics[width=2.0in,angle=0]{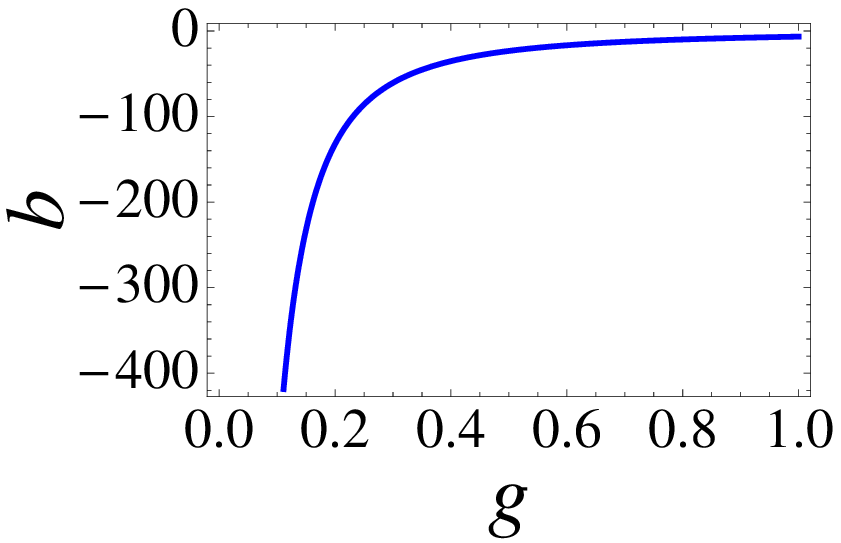}
\includegraphics[width=2.0in,angle=0]{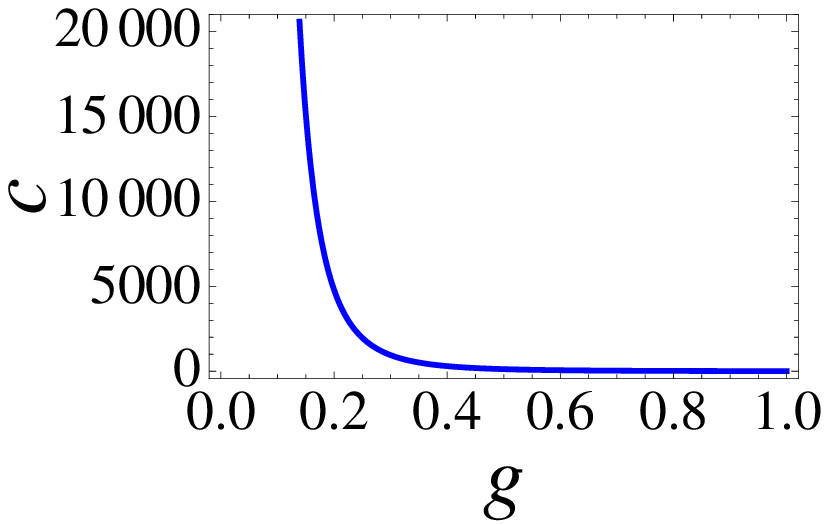}
\caption{Plots of the GL coefficients $a,b$ and $c$ vs $g$ for $U=2.0t$. One can see that $a$ changes sign as $g$ increases while $b < 0$ and $c > 0$ for all values of $g$.}
\label{GL}
\end{center}
\end{figure}
Here $E^{\prime\prime}$ is second derivative of $E_{gnd}$ and so on.
To decide about the nature of the phase transition, it is sufficient to look at the signs of the coefficients $a,b$ and $c$~\cite{Huang}. For $b,c >0$, if $a>0$, $m_s=0$ is the only point of minima of the ground state energy. As $a$ changes sign, the system undergoes a second order phase transition to the magnetically ordered phase with $m_s^2 = \f{1}{2c}(-b+\sqrt{b^2-4ac})$.  For $c >0$ and $b<0$, we have a first order phase transition at $b=-4\sqrt{ca/3}$ where the magnetisation $m_s$ changes discontinuously by the amount $\lbr \f{3a}{c}\rbr^{1/4}$. 

Expressions for the GL coefficients in the Taylor series expansion of the ground state energy in Eq.~\ref{GLE} are given below,\\
$E_0 = -2\sum_k E(\epsilon_k)  -\f{U}{2}\delta n^2$ \\
$\f{a}{2} = U+2(\f{U}{2})^2 \sum_k\f{1}{E(\epsilon_k)} [r^2-1]$ \\
$\f{b}{4} = 6(\f{U}{2})^4\sum_k\f{1}{[E(\epsilon_k)]^3}[1-6r^2+5r^4]$ \\
$\f{c}{6} = 90(\f{U}{2})^6\sum_k\f{1}{[E(\epsilon_k)]^5}\lbs -1+15r^2-35r^4+21r^6\rbs $
\\
Here $E(\epsilon_k) = E_{\sigma}(\epsilon_k)|_{m_s=0} = \sqrt{\epsilon_k^2+g^2}$ with $g=g_\sigma|_{m_s=0}=\Delta-\frac{U}{2}\delta n$ and $r=\frac{g}{E(\epsilon_k)}$.

We have numerically calculated the coefficients $a,b$ and $c$ and found that for all values of $\Delta$ and $U/t$ studied, $c$ is always positive while $b$ is always negative. $a>0$ for $U <U_1$ and becomes negative for $U >U_1$ where the value of $U_1$ depends upon $\Delta/t$. 
For the Bethe lattice of infinite connectivity, the integrals involved in the above equations can be done analytically and we get the following expressions for the GL coefficients:
\\
\\
$\frac{a}{2}=U+\frac{(U/2)^2}{\pi t^2}\lbs 4g \mathcal{E}(-\frac{4t^2}{g^2})-4\frac{(2t^2+g^2)}{g}\mathcal{K}(-\frac{4t^2}{g^2})\rbs$ 
\bea
\frac{b}{4}=\frac{8(U/2)^4}{\pi t^2 g(4t^2+g^2)}\lbs (g^2-4t^2)\mathcal{E}(-\frac{4t^2}{g^2})-(g^2+4t^2)\mathcal{K}(-\frac{4t^2}{g^2})\rbs \nonumber
\eea
\be
\frac{c}{6}=\frac{96(U/2)^6}{\pi t^2g^3(4t^2+g^2)^4}\lbs c1(g)\mathcal{E}(-\frac{4t^2}{g^2})-c2(g)\mathcal{K}(-\frac{4t^2}{g^2})\rbs
\label{GLC}
\ee
with $c1(g)=32+32g^2+18g^4-g^6$ and $c2(g)=16+24g^2+g^4-g^6$. 
Here $\mathcal{K}(x)= \int_0^{\pi/2} [1-x sin^2(\theta)]^{-1/2} d~\theta$ is the complete elliptic integral of the first kind and $\mathcal{E}(x)=\int_0^{\pi/2} [1-x sin^2(\theta)]^{1/2} d~\theta$ is the complete elliptic integral
 of the second kind.  Fig.~\ref{GL} shows the plots of GL coefficients $a,b$ and $c$ (obtained from Eq.~\ref{GLC}) vs $g$ for a fixed value of $U$. As $g\rightarrow 0$, $\mathcal{K}(-\frac{4t^2}{g^2}) \rightarrow 0$ while $\mathcal{E}(-\frac{4t^2}{g^2})\rightarrow \infty$.  Thus for $g < 2t$, which is the regime of interest, $b$ is always negative. 
Thus following the GL approach~\cite{Huang} we conclude that the transition from the PM BI to the AFM phase in the half filled IHM is always of first order in nature, even for very small values of $\Delta/t$.

\end{document}